\documentclass[twocolumn,showpacs,aps,superscriptaddress]{revtex4-1}%
\usepackage{amsmath}
\usepackage{amssymb}
\usepackage{amsfonts}
\usepackage{amssymb}
\usepackage{graphicx}
\usepackage{txfonts}
\usepackage[linkcolor=blue,anchorcolor=black,citecolor=blue]{hyperref}%
\providecommand{\U}[1]{\protect\rule{.1in}{.1in}}

\begin{document}
\title{Quantum sensing enhanced by adaptive periodic quantum control}
\author{Yi-Nan Fang}
\affiliation{Beijing Computational Science Research Center, Beijing 100193, China}
\affiliation{Synergetic Innovation Center of Quantum Information and Quantum Physics,
University of Science and Technology of China, Hefei, Anhui 230026, China}
\affiliation{State Key Laboratory of Theoretical Physics, Institute of Theoretical Physics,
Chinese Academy of Sciences, and University of the Chinese Academy of
Sciences, Beijing 100190, China}
\author{Xing Xiao}
\affiliation{Beijing Computational Science Research Center, Beijing 100193, China}
\author{Chang-Pu Sun}
\affiliation{Beijing Computational Science Research Center, Beijing 100193, China}
\affiliation{Synergetic Innovation Center of Quantum Information and Quantum Physics,
University of Science and Technology of China, Hefei, Anhui 230026, China}
\author{Wen Yang}
\email{wenyang@csrc.ac.cn}
\affiliation{Beijing Computational Science Research Center, Beijing 100193, China}
\author{Nan Zhao}
\email{nzhao@csrc.ac.cn}
\affiliation{Beijing Computational Science Research Center, Beijing 100193, China}
\affiliation{Synergetic Innovation Center of Quantum Information and Quantum Physics,
University of Science and Technology of China, Hefei, Anhui 230026, China}

\begin{abstract}
Using a single quantum probe to sense other quantum objects offers distinct
advantages but suffers from some limitations that may degrade the sensing
precision severely, especially when the probe-target coupling is weak. Here we
propose a strategy to improve the sensing precision by using the quantum probe
to engineer the evolution of the target. We consider an exactly solvable
model, in which a qubit is used as the probe to sense the frequency of a
harmonic oscillator. We show that by applying adaptive periodic quantum
control on the qubit, the sensing precision can be enhanced from $1/T$ scaling
with the total time cost $T$ to $1/T^{2}$ scaling, thus improving the
precision by several orders of magnitudes. Such improvement can be achieved
without any direct access to the oscillator and the improvement increases with
decreasing probe-target coupling. This provides a useful routine to
ultrasensitive quantum sensing of weakly coupled quantum objects.

\end{abstract}

\maketitle

Using single quantum objects as quantum probes for sensing provides distinct
advantages, e.g., high spatial resolution
\cite{TaylorNatPhys2008,BalasubramanianNature2008,MazeNature2008},
integrability, and miniature of devices, in comparison with macroscopic
probes. Due to recent experimental progress in controlling single quantum
objects, such as single trapped ions, superconducting qubits, and single
defect spins in solids
\cite{RobledoNature2011,ToganNature2010,ToganNature2011,MazeNJP2011,OBrienPRB2001,MorelloNature2010}%
, atomic scale sensing with single quantum probes is now made possible
\cite{DegenRMP2017}, and may trigger new applications in broad fields
including chemistry, biology and material sciences. However, the widely used
quantum resources -- large-scale entanglement and interactions among different
quantum probes -- are no longer available for a single quantum probe.
Moreover, a large family of tasks requires sensing quantum objects weakly
coupled to the quantum probe, where direct access (e.g., initialization,
manipulation, or measurement) to the target quantum object is not available.
These limitations may severely degrade the key figure of merit -- the sensing
precision. It is important to identify and utilize available resources to
improve the sensing precision of single quantum probes for weakly coupled
quantum objects.

The coherent evolution time $T$ is an important quantum resource. Previous
works
\cite{ZhaoNatNano2011,ZhaoNatNano2012,TaminiauPRL2012,KolkowitzPRL2012,CaiNJP2013,LondonPRL2013,LaraouiNatCommun2013,ShiNatPhys2014,LangPRX2015,BossPRL2016,ZaiserNC2016,MaPRA2016,MaPRA2016a,ShuPRA2017,LiuPRL2017a}
on sensing quantum objects mostly use non-adaptive schemes and their precision
is upper bounded by a $1/T$ time scaling. By contrast, for sensing classical
signals, recent theoretical works show that using adaptive techniques allows
universal $1/T$ scaling \cite{YuanPRL2015,YuanPRL2016} (and $1/T^{2}$ scaling
for special tasks \cite{PangNC2017}), consistent with available experimental
reports \cite{NusranNatNano2012,WaldherrNatNano2012,BonatoNatNano2016}.
However, they are not applicable to sensing weakly coupled quantum objects due
to the lack of direct access to the target. Remarkably, a recent breakthrough
improves the time scaling to $1/T^{3/2}$ by using the continuous sampling
technqiue \cite{SchmittScience2017,BossScience2017}.

In this work, we propose a strategy for improving the precision for sensing
weakly coupled quantum objects. The key is to combine periodic quantum control
on the quantum probe
\cite{ZhaoNatNano2011,ZhaoNatNano2012,TaminiauPRL2012,KolkowitzPRL2012,CaiNJP2013,LondonPRL2013,LaraouiNatCommun2013,ShiNatPhys2014,LangPRX2015,BossPRL2016,ZaiserNC2016,MaPRA2016,MaPRA2016a}
with adaptive techniques to steer the evolution of the target for maximal
information flow from the target to the probe. We consider a single qubit as a
quantum probe to estimate the frequency $\omega$ of a harmonic oscillator -- a
paradigmatic hybrid system that has attracted a lot of interest recently
\cite{AspelmeyerRMP2014}. We show that applying adaptive\textit{ }periodic
control on the qubit improves the time scaling of the precision from $1/T$ to
$1/T^{2}$, thus enhancing the precision by several orders of magnitudes.
Interestingly, this improvement can be achieved \textit{without} any direct
access to the oscillator, and the improvement increases with decreasing
coupling strength between the qubit and the oscillator. This study highlights
adaptive periodic quantum control as a useful route for ultra-sensitive
quantum sensing of weakly coupled quantum objects.

\section{Results}

\subsection{Sensing other quantum objects: limitations and opportunities}

A typical protocol to estimate an unknown parameter $\theta$ with a quantum
system consists of three steps: (1) The system starts from certain initial
state $\hat{\rho}$ and undergoes certain $\theta$-dependent evolution into the
final state $\hat{\rho}_{\theta}$. The information in $\hat{\rho}_{\theta}$ is
quantified by the quantum Fisher information $\mathcal{F}$
\cite{BraunsteinPRL1994}. (2) A measurement on $\hat{\rho}_{\theta}$ gives an
outcome randomly sampled from all possible outcomes $\{x_{m}\}$ according to
certain measurement distribution $P(x_{m}|\theta)$ conditioned on $\theta$.
The information in each outcome is quantified by the classical Fisher
information $F=\sum_{m}P(x_{m}|\theta)[\partial_{\theta}\ln P(x_{m}%
|\theta)]^{2}$ \cite{KayBook1993}, which obeys $F\leq\mathcal{F}$. (3) Steps
(1) and (2) are repeated $\nu$ times and the $\nu$ outcomes are processed to
yield an estimator $\theta_{\mathrm{est}}$ to $\theta$. The precision of
$\theta_{\mathrm{est}}$ is quantified by its statistical error $\delta\theta$.

For unbiased estimators \cite{KayBook1993}, the precision $\delta\theta$ is
fundamentally limited by the Cram\'{e}r-Rao bound
\cite{HelstromBook1976,BraunsteinPRL1994}%
\begin{equation}
\delta\theta\geq\frac{1}{\sqrt{\nu F}}\geq\frac{1}{\sqrt{\nu\mathcal{F}}}.
\label{CRB}%
\end{equation}
For optimal performance, optimal initial state and evolution should be used to
maximize $\mathcal{F}$, optimal measurements should be designed to make
$F=\mathcal{F}$, and optimal unbiased estimators should be used to saturate
the first inequality of Eq. (\ref{CRB}).

Sensing classical signals amounts to estimating certain parameter of the
quantum probe. The simplest example is to estimate a real parameter $\theta$
in the probe Hamiltonian $\theta\hat{H}$. Starting from an initial state
$|\psi\rangle$, the quantum probe evolves for an interval $T$ into a final
state $|\psi_{\theta}\rangle\equiv e^{-i\theta T\hat{H}}|\psi\rangle$ with
$\mathcal{F}=4H_{\mathrm{rms}}^{2}T^{2}$, where $H_{\mathrm{rms}}%
\equiv(\langle\psi|\hat{H}^{2}|\psi\rangle-\langle\psi|\hat{H}|\psi\rangle
^{2})^{1/2}$ is the\ fluctuation of $\hat{H}$ in the initial state. If the
subsequent measurement and estimator are both optimized to avoid information
loss, then after $\nu$ repeated measurements, Eq. (\ref{CRB}) gives
\begin{equation}
\delta\theta=\frac{1}{2H_{\mathrm{rms}}\sqrt{\nu}T}. \label{DW0}%
\end{equation}
The precision improves with $\nu$ according to the classical scaling
$1/\sqrt{\nu}$, but improves with $T$ according to the enhanced scaling $1/T$
due to the linear phase accumulation $e^{-i\theta T\hat{H}}$
\cite{GiovannettiPRL2006}.

Sensing other quantum objects amounts to estimating certain parameter of the
target quantum object. In this case, the lack of direct access to the target
may degrade the sensing precision significantly: (i) The lack of
initialization and direct control over the target may degrade $\mathcal{F}$ in
the final state; (ii)\ The lack of direct measurement over the target may
cause information loss during the conversion from $\mathcal{F}$ to $F$; (iii)
The unintended evolution of the target due to the backaction of the quantum
probe may further degrade $\mathcal{F}$. To illustrate (iii), we consider
using a quantum probe with Hamiltonian $\hat{H}_{\mathrm{p}}$ to estimate a
real parameter $\theta$ in the target Hamiltonian $\theta\hat{H}$ through the
probe-target coupling $\hat{V}$. The coupled system starts from $|\Psi\rangle$
and evolves under the total Hamiltonian $\mathcal{H}=\theta\hat{H}+\hat
{H}_{\mathrm{p}}+\hat{V}$ for an interval $T$ into the final state
$e^{-i\mathcal{H}T}|\Psi\rangle$ with $\mathcal{F}=4\bar{H}_{\mathrm{rms}}%
^{2}T^{2}$, where $\bar{H}_{\mathrm{rms}}$ is the fluctuation of $\bar
{H}\equiv(1/T)\int_{0}^{T}\hat{H}(t)dt$ in the initial state
\cite{PangPRA2014,LiuSciRep2015} and $\hat{H}(t)\equiv e^{i\mathcal{H}t}%
\hat{H}e^{-i\mathcal{H}t}$ undergoes unintended evolution when $[\hat{V}%
,\hat{H}]\neq0$. If $\hat{H}$ is off-diagonal in the eigenbasis of
$\mathcal{H}$, then $\bar{H}\propto1/T$ at large $T$, so increasing $T$ does
not improve the precision at all.

Fortunately, in addition to causing unintended evolution of the target, the
backaction of the quantum probe can also be utilized to steer the evolution of
the target Hamiltonian $\hat{H}(t)$ by appropriate quantum control over the
probe. This provides an opportunity to improve the precision for sensing other
quantum objects.

\subsection{Quantum sensing by periodic quantum control}

\begin{figure}[ptb]
\begin{centering}
\includegraphics[width=\columnwidth]{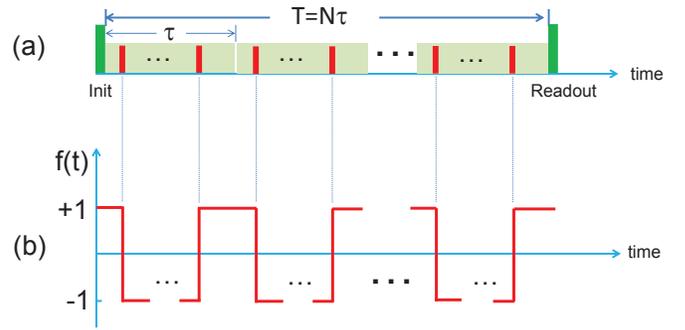}
\par\end{centering}\caption{\textbf{Quantum sensing by adaptive periodic
quantum control.} (a) The qubit is first initialized into the $\hat{\sigma
}_{x}=+1$ eigenstate $|+\rangle$, and then experiences a periodic quantum
control with $N$ identical control units of duration $\tau$. Each control unit
consists of an even number of instantaneous $\pi$ pulses. Finally $\hat
{\sigma}_{x}$ is readout by a projective measurement. (b) Modulation function
of this periodic quantum control. The vertical dashed lines are guides to the
eye.}%
\label{G_PROTOCOL}%
\end{figure}

We consider using a qubit as a quantum probe to sense the frequency $\omega$
of a harmonic oscillator that cannot be accessed directly. The Hamiltonian is
the sum of the qubit term $\omega_{0}\hat{\sigma}_{z}/2$, the oscillator term
$\omega\hat{b}^{\dagger}\hat{b}$, and the qubit-oscillator coupling
$(\lambda/2)(\hat{b}^{\dagger}+\hat{b})\hat{\sigma}_{z}$
\cite{NeukirchOL2013,ScalaPRL2013,ZhaoPRA2014}, where $\hat{\sigma}_{x,y,z}$
are Pauli matrices for the qubit. This model has been realized experimentally
in various hybrid quantum systems
\cite{LaHayeNature2009,HungerPRL2010,BennettPRL2010,ArcizetNatPhys2011,KolkowitzS2012,YeoINatNano2014}
by coupling a two-level system to a mechanical nano-oscillator
\cite{AspelmeyerRMP2014}. As shown in Fig. \ref{G_PROTOCOL}(a), the quantum
control on the qubit consists of $N$ identical units of duration $\tau$ and
each unit consists of an even number of $\pi$-pulses. Each $\pi$-pulse causes
an instantaneous $\pi$-rotation $e^{-i(\pi/2)\hat{\sigma}_{x}}$ of the qubit
around the $x$ axis. In the interaction picture of the qubit, the total
Hamiltonian is
\[
\mathcal{H}(t)=\omega\hat{b}^{\dagger}\hat{b}+f(t)\frac{\lambda}{2}(\hat
{b}^{\dagger}+\hat{b})\hat{\sigma}_{z},
\]
where $f(t)$ is the modulation function associated with the quantum control
\cite{YangRPP2017}:\ it starts from $f(0)=+1$ and changes its sign at the
timings of each $\pi$-pulse [Fig. \ref{G_PROTOCOL}(b)]. Using the Wei-Norman
algebra method \cite{SunCTP1991}, the evolution operator during the total
period $T\equiv N\tau$ of the quantum control is obtained as $\hat
{U}=e^{-i\omega T\hat{b}^{\dagger}\hat{b}}\hat{D}(\hat{\sigma}_{z}\alpha)$,
where $\hat{D}(z)=e^{z\hat{b}^{\dagger}-z^{\ast}\hat{b}}$ is the oscillator
displacement operator and
\[
\alpha=-i\frac{\lambda}{2}\int_{0}^{N\tau}f(t)e^{i\omega t}dt=\alpha_{1}K,
\]
with $\alpha_{1}\equiv\alpha|_{N=1}$ for a single control unit and%
\[
K=\sum_{n=0}^{N-1}e^{i\omega n\tau}=\frac{e^{i\omega N\tau/2}}{e^{i\omega
\tau/2}}\frac{\sin\frac{\omega N\tau}{2}}{\sin\frac{\omega\tau}{2}}%
\]
for the interference from $N$ control units.

Before the quantum control, we initialize the qubit into the $\hat{\sigma}%
_{x}=+1$ eigenstate $|+\rangle=(|\uparrow\rangle+|\downarrow\rangle)/\sqrt{2}%
$, but leave the oscillator in an arbitrary initial state $\hat{\rho}$ since
the oscillator cannot be initialized. The evolution $\hat{U}$ during the
quantum control drives the coupled system into an entangled final state
$\hat{U}|+\rangle\langle+|\hat{\rho}\hat{U}^{\dagger}$. Since the oscillator
cannot be measured, only the quantum Fisher information $\mathcal{F}$
contained in the reduced density matrix of the qubit, $\hat{\rho}_{\mathrm{p}%
}\equiv1/2+\left(  L|\uparrow\rangle\langle\downarrow|+h.c.\right)  /2$, can
be converted into the classical Fisher information $F$, where $L\equiv
\langle\hat{D}(2\alpha)\rangle$ is the off-diagonal coherence of the qubit and
$\left\langle \cdots\right\rangle \equiv\operatorname*{Tr}\hat{\rho}(\cdots)$
denotes the average over the initial state of the oscillator. Here we assume
$\hat{\rho}$ commutes with $\hat{b}^{\dagger}\hat{b}$ and leave the
generalization to an arbitrary $\hat{\rho}$ to the next section. In this case,
$L$ is real and $\mathcal{F}=(\partial_{\omega}L)^{2}/(1-L^{2})$
\cite{ZhongPRA2013}. Let $\bar{n}\equiv\langle\hat{b}^{\dagger}\hat{b}\rangle
$, when
\begin{equation}
\sqrt{2\bar{n}+1}\left\vert \alpha\right\vert \ll1, \label{CD}%
\end{equation}
we obtain $L\approx1-2(2\bar{n}+1)|\alpha|^{2}$ and hence $\mathcal{F}%
\approx4(2\bar{n}+1)(\partial_{\omega}|\alpha|)^{2}$. At the end of the
quantum control, a projective measurement of $\hat{\sigma}_{x}$ on the qubit
yields an outcome randomly sampled from $\{+1,-1\}$ according to the
probability $P(\pm1|\omega)=(1\pm L)/2$ and the classical Fisher information
contained in each outcome is obtained as $F=\mathcal{F}$. Such measurements
are experimentally available in traditional nuclear magnetic resonance and
electron spin resonance systems. The ultimate sensing precision follows from
Eq. (\ref{CRB}) as
\begin{equation}
\delta\omega=\frac{1}{\sqrt{\mathcal{F}}}=\frac{1}{2\sqrt{2\bar{n}%
+1}\left\vert (\partial_{\omega}|\alpha|)\right\vert }. \label{DW1}%
\end{equation}

\begin{figure}[ptb]
\begin{centering}
\includegraphics[width=\columnwidth]{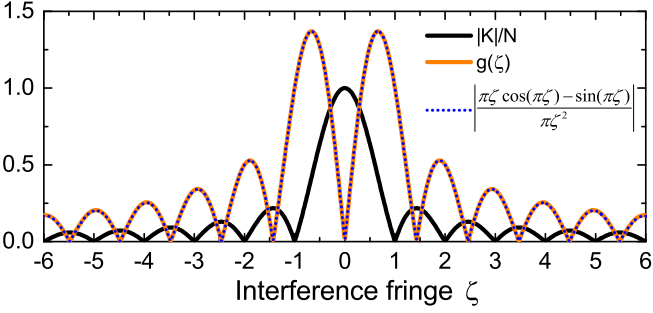}
\par\end{centering}\caption{\textbf{Interference fringes from periodic quantum
control with }$N=50$\textbf{ control units}. Here $\zeta\equiv N(\omega
\tau/2\pi-1)$ labels the interference fringes near the major peak at
$\omega\tau=2\pi$.}%
\label{G_FRINGE}%
\end{figure}

For large $N$, $|K|$ and hence $|\alpha|$ as functions of $\omega\tau$ exhibit
many interference fringes with major peaks at integer multiples of $2\pi$. We
focus on the major peak at $2\pi$ and label the surrounding interference
fringes by $\zeta\equiv N\left(  \omega\tau/2\pi-1\right)  $, e.g., $\zeta=0$
labels the major peak and $\zeta=\pm1,\pm2$ labels the nodes (see the black
solid line in Fig. \ref{G_FRINGE}). For $\left\vert \zeta\right\vert \ll N$,
$\left\vert \alpha_{1}\right\vert $ is nearly a constant, so Eq. (\ref{DW1})
simplifies to%
\begin{equation}
\delta\omega\approx\frac{\pi}{g\tilde{\lambda}T^{2}}, \label{DW}%
\end{equation}
where $\tilde{\lambda}\equiv\sqrt{2\bar{n}+1}\left\vert \alpha_{1}\right\vert
/\tau$ is nearly a constant and $g(\zeta)\equiv|(\partial_{\zeta}|K|/N)|$
approaches a universal function (see Fig. \ref{G_FRINGE})
\[
g(\zeta)\approx\left\vert \frac{\pi\zeta\cos(\pi\zeta)-\sin(\pi\zeta)}%
{\pi\zeta^{2}}\right\vert \approx\left\{
\begin{array}
[c]{l}%
\frac{\pi^{2}\left\vert \zeta\right\vert }{3}e^{-(\pi\zeta)^{2}/10}%
\ (|\zeta|\ll1),\\
\\
\frac{|\cos(\pi\zeta)|}{\left\vert \zeta\right\vert }\ \ \ \ \ (|\zeta
|\gtrsim1).
\end{array}
\right.
\]
There are two tunable parameters: the duration $\tau$ of each control unit and
the total number $N$ of control units. We set $\tau$ close to $2\pi/\omega$ to
make $\zeta\approx1$, so that $\left\vert \alpha\right\vert $ is sufficiently
small to satisfy Eq. (\ref{CD}), while $g\approx1$ is large to optimize the
sensing precision. With $\tau$ largely fixed, we can increase the evolution
time $T=N\tau$ by increasing $N$, so $\delta\omega\approx1/(\tilde{\lambda
}T^{2})$. Interestingly, this $1/T^{2}$ scaling originates from the
interference between different control units: $\partial_{\omega}|K|\propto
T^{2}$, while the internal structure of each control unit only affects the
value of $\left\vert \alpha_{1}\right\vert /\tau$ and hence $\tilde{\lambda}$,
e.g., $\left\vert \alpha_{1}\right\vert /\tau\approx\lambda/\pi$ for the
Carr--Purcell--Meiboom--Gill sequence \cite{CarrPR1954,MeiboomRSI1958} with
two $\pi$-pulses locate at $\tau/4$ and $3\tau/4$ in one control unit.

\subsection{Origin of $1/T^{2}$ scaling}

Compared with the previous work \cite{YuanPRL2015,YuanPRL2016} for sensing
classical signals, where sophisticated feedback control are required to
achieve the universal $1/T$ scaling, it is interesting that for the more
challenging task -- sensing quantum objects, our protocol can achieve the
$1/T^{2}$ scaling by applying a simple periodic quantum control on the qubit
\textit{without }any direct access to the oscillator. The solution is that the
previous derivation of the universal $1/T$ \cite{YuanPRL2015,YuanPRL2016}
scaling for sensing classical signals assumes the quantum probe has a fixed
and bounded spectrum. When this restriction is lifted, e.g., if the
Hamiltonian itself increases with time $t$ as $\theta t^{k}\hat{H}$, then the
precision $\delta\theta$ would be given by Eq. (\ref{DW0}) with
$H_{\mathrm{rms}}\rightarrow T^{k}\hat{H}_{\mathrm{rms}}$, i.e., $\delta
\theta\propto1/T^{k+1}$ \cite{PangNC2017}. By contrast, although sensing
quantum objects suffers from the lack of direct access to the target, the
spectrum of the target may be unbounded (even though the spectrum of the probe
is bounded) and can further be manipulated indirectly via the probe, so the
time scaling is not limited to $1/T$, but instead can be raised by engineering
the evolution of the target, e.g., through the periodic driving on the qubit
in our qubit-oscillator model. However, the lack of direct access to the
target does lead to some surprising consequences, as we discuss now.

First, the condition Eq. (\ref{CD}) for achieving the $1/T^{2}$ scaling leads
to $\hat{U}\approx e^{-i\omega T\hat{b}^{\dagger}\hat{b}}$, i.e., the final
state of the coupled system at the end of the quantum control should largely
coincide with their initial product state $|+\rangle\langle+|\otimes\hat{\rho
}$. In other words, achieving the $1/T^{2}$ scaling requires \textit{neither}
appreciable probe-target entanglement \textit{nor} appreciable energy
fluctuation in the initial or final state of the coupled system, despite a
large amount of energy exchange during the evolution. For example, even if the
oscillator starts from (and ends up with)\ the lowest-energy vacuum state, we
still obtain Eq. (\ref{DW}) (albeit with $\bar{n}=0$). This differs from
sensing classical signals, where large scale entanglement and large energy
fluctuation in the initial or final state are standard quantum resources to
improve the precision \cite{GiovannettiPRL2006}, e.g., according to Eq.
(\ref{DW0}), to achieve optimal precision, the quantum system should start
from (and end with) a highly excited state -- an equal superposition of the
highest eigenstate and the lowest eigenstate of $\hat{H}$.

Second, if we tune $\tau$ to make $|\alpha|\gg1$, then the evolution $\hat
{U}=e^{-i\omega T\hat{b}^{\dagger}\hat{b}}\hat{D}(\hat{\sigma}_{z}\alpha)$
would lead to large bifurcated displacement of the oscillator by $\pm\alpha$
for the qubit state being $|\uparrow\rangle$ or $|\downarrow\rangle$, so the
final state of the coupled system is highly entangled. Although this state do
contain a lot of quantum Fisher information about $\omega$, converting all of
them into classical Fisher information would require projective measurements
in the qubit-oscillator entangled basis, which is unavailable. The only object
that can be measured is the final state of the qubit which, for $|\alpha|\gg
1$, is almost completely random and contains little quantum Fisher information
about $\omega$. In other words, feeding a large amount of energies into the
final state of the oscillator degrades, instead of improves, the sensing precision.

Third, thermal fluctuation of the oscillator usually degrades the sensing
precision dramatically, e.g., if the initial state of the oscillator is a
thermal state, then using the Linked-cluster expansion \cite{YangRPP2017}
gives $L=e^{-2(2\bar{n}+1)|\alpha|^{2}}$, so $\mathcal{F}\sim(\partial
_{\omega}L)^{2}$ is exponentially suppressed when $\sqrt{2\bar{n}+1}%
|\alpha|\gg1$, similar to the case of measuring the frequency of a harmonic
oscillator under classical driving by directly monitoring its positions.
However, in our protocol, we can tune $\tau$ to make $|\alpha|$ sufficiently
small so that Eq. (\ref{CD}) is satisfied, then $\mathcal{F}\propto\bar{n}$
and the sensing precision $\delta\omega\propto1/\sqrt{\bar{n}}$ improves with
$\bar{n}$ \cite{ZhaoPRA2014}.

In deriving Eqs. (\ref{DW1}) and (\ref{DW}), we have assumed $[\hat{\rho}%
,\hat{b}^{\dagger}\hat{b}]=0$ to make $L$ a real number. When this constraint
is lifted, $L$ is in general complex, so $\mathcal{F}=|\partial_{\omega}%
L|^{2}+|L|^{2}(\partial_{\omega}|L|)^{2}/(1-|L|^{2})$ \cite{ZhongPRA2013},
where $L\approx1+4i\operatorname{Im}\alpha\langle\hat{b}^{\dagger}%
\rangle+4\operatorname{Re}\alpha^{2}\langle b^{\dagger2}\rangle-2(2\bar
{n}+1)|\alpha|^{2}$ for small $|\alpha|$. As long as $\zeta=O(1)$, both
$\partial_{\omega}\alpha$ and $\partial_{\omega}|\alpha|$ are of the order
$T^{2}$, so we expect $\partial_{\omega}L,\partial_{\omega}|L|=O(T^{2})$ and
$\mathcal{F}=O(T^{4})$, i.e., the $1/T^{2}$ scaling holds for a general
oscillator initial state.

\subsection{Adaptive quantum control}

\begin{figure}[ptb]
\begin{centering}
\includegraphics[width=\columnwidth]{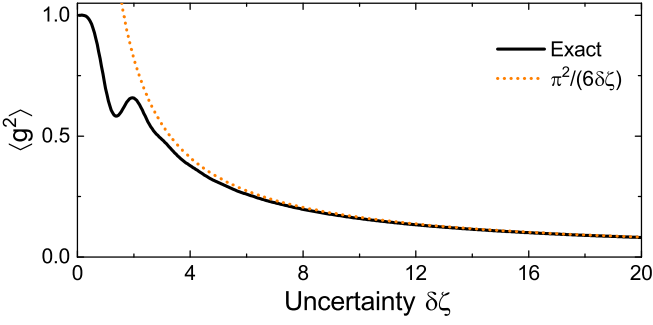}
\par\end{centering}\caption{\textbf{Performance of our protocol vs.
uncertainty in tuning the interference fringes.}}%
\label{G_G2}%
\end{figure}

According to Eq. (\ref{DW}), the $1/T^{2}$ scaling can be achieved in two
steps. First, we should tune $\tau$ to make $\zeta$ locate at the first node
$\zeta=1$, so that $\left\vert \alpha\right\vert =0$ satisfies Eq. (\ref{CD})
and $g=1$. Second, we should increase $N$ to increase the total time $T$, so
that $\delta\omega\approx\pi/(\tilde{\lambda}T^{2})$. However, our limited
prior knowledge about $\omega$ -- the unknown parameter -- makes it impossible
to make $\zeta$ locate at the first node precisely. If we our knowledge about
$\omega$ has an uncertainty $\delta\omega$, then we would suffer from an
uncertainty $\delta\zeta\equiv N\tau\delta\omega/2\pi$ in tuning the value of
$\zeta$, so the achievable sensing precision is roughly given by Eq.
(\ref{DW}) with $g(\zeta)$ replaced by $\langle g^{2}\rangle^{1/2}$, where
$\langle g^{2}\rangle$ is the average of $g^{2}(\zeta)$ over the region
$[1-\delta\zeta,1+\delta\zeta]$. As shown in Fig. \ref{G_G2}, $\langle
g^{2}\rangle\propto1/\delta\zeta$ at large $\delta\zeta$, thus if
$\delta\omega$ is fixed, then $\langle g^{2}\rangle\propto1/N$ leads to
$1/T^{3/2}$ scaling according to Eq. (\ref{DW}). To achieve the $1/T^{2}$
scaling, we need to ensure $\delta\zeta\lesssim1$ and Eq. (\ref{CD})
simultaneously. This limits the $1/T^{2}$ scaling to
\begin{equation}
T\lesssim T_{\max}\equiv\sqrt{\frac{2\pi}{\delta\omega\max\{\delta
\omega,\tilde{\lambda}\}}},\label{CONSTRAINT}%
\end{equation}
as determined by $\delta\omega$. This limitation can be lifted by using
adaptive techniques.

\begin{figure}[ptb]
\begin{centering}
\includegraphics[width=\columnwidth]{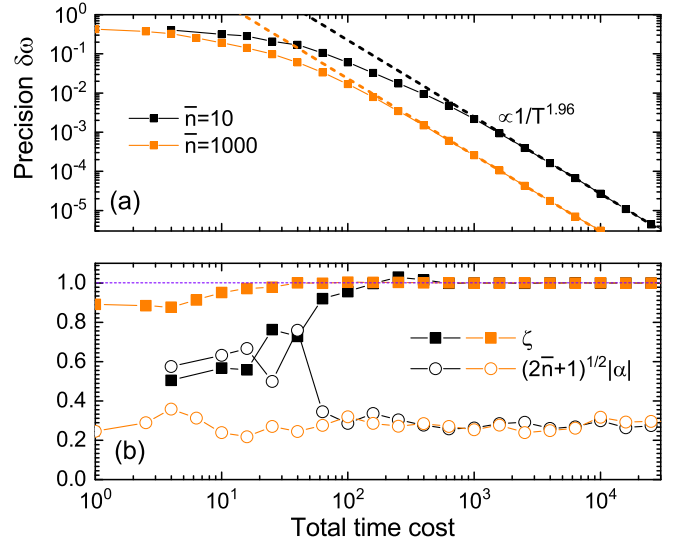}
\par\end{centering}\caption{\textbf{Numerical simulation of our protocol for a
thermal initial state of the oscillator}. (a) Sensing precision $\delta\omega$
vs. total time cost. (b) Evolution of $\zeta$ (squares) and $\sqrt{2\bar{n}%
+1}|\alpha|$ (circles) during the adaptive measurement. Each data is obtained
by averaging the results of 500 repeated simulations. The dashed lines are
linear fits to the simulation data. The true value of the frequency
$\omega=50$, the thermal population $\bar{n}=10$ (black squares and circles)
or $1000$ (orange squares and circles), the coupling strength $\lambda=0.1$,
and the prior knowledge $\delta\omega_{0}=0.5$, and $\omega_{0}=50.5.$}%
\label{G_SIMULATION}%
\end{figure}

Suppose before the sensing, we have an unbiased estimator $\omega_{0}$ with
uncertainty $\tilde{\lambda}<\delta\omega_{0}\ll\omega_{0}$. This prior
knowledge may come from preliminary measurements without quantum control. The
entire scheme consists of many adaptive steps. The key idea is to utilize the
knowledge acquired from the measurements in every step to reduce the
uncertainty $\delta\omega$ in our knowledge about $\omega$, so that a longer
evolution time $T$ can be used in the next step according to Eq.
(\ref{CONSTRAINT}) (see Methods for details). In Fig. \ref{G_SIMULATION}, we
show the results from our numerical simulation for a thermal initial state of
the oscillator. Figure \ref{G_SIMULATION}(a) shows that (i) after a few tens
of adaptive steps, the sensing precision begins to improve with the total time
cost $\mathbb{T}$ according to the $1/\mathbb{T}^{2}$ scaling, where
$\mathbb{T}$ can be extended indefinitely by increasing the number of adaptive
steps; (ii)\ increasing the thermal fluctuation from $\bar{n}=10$ to $\bar
{n}=1000$ improves the precision significantly. The onset of the
$1/\mathbb{T}^{2}$ scaling can be understood from Fig. \ref{G_SIMULATION}%
(b):\ after a few tens of adaptive steps, the value of $\zeta$ is tuned
accurately to the first node $\zeta=1$ and Eq. (\ref{CD}) is well-satisfied.

\section{Discussions}

To quantify the effect of the adaptive quantum control, we compare the sensing
precision $\delta\omega\sim\pi/(\tilde{\lambda}\mathbb{T}^{2})$ under the
quantum control to that without any control. The latter corresponds to
$f(t)\equiv1$ and hence $\left\vert \alpha\right\vert =(\lambda/\omega
)\left\vert \sin(\omega\mathbb{T}/2)\right\vert $ for an evolution time
$\mathbb{T}$, so the precision follows from Eq. (\ref{DW1}) as $\delta
\omega_{\mathrm{free}}\sim\omega/(\tilde{\lambda}\mathbb{T})$. Therefore,
improving the precision from $\tilde{\lambda}$ to $\tilde{\lambda}/K$ requires
a time cost $\mathbb{T}_{\mathrm{free}}\sim K\omega/\tilde{\lambda}^{2}$
without any control or $\mathbb{T}\sim\sqrt{K}/\tilde{\lambda}$ under the
quantum control, i.e., the quantum control reduces the time cost by a factor
\[
\frac{\mathbb{T}_{\mathrm{free}}}{\mathbb{T}}\sim\sqrt{K}\frac{\omega}%
{\tilde{\lambda}}\sim\sqrt{K}\frac{\omega}{\lambda}%
\]
that increases with increasing desired precision (i.e., increasing $K$) and
decreasing coupling strength $\lambda$. In other words, our protocol is
especially suited to high-precision sensing of remote quantum objects that are
weakly coupled to the quantum probe -- a most important yet challenging task.

In practice, the evolution time $T$ would be ultimately limited by the finite
coherence time $T_{2}$ of the qubit \footnote{Here we consider using the
quantum probe for high-precision sensing of a well-defined frequency of the
target quantum object. This requires the coherence time $T_{\mathrm{tar}}$ of
the target to be much longer than that of the quantum probe, otherwise the
target frequency would be broadened by $\sim1/T_{\mathrm{tar}}$, making
high-precision sensing impossible.}, so the coherent evolution time $T=N\tau$
in each measurement would reach $T_{2}$ after some adaptive steps. Afterwards,
the optimal strategy is to repeat the measurements with evolution time $T\sim
T_{2}$ in all the subsequent steps, so the performance is quantified by the
frequency sensitivity $\mathbb{S}\equiv\delta\omega\sqrt{T}$, which is
$\mathbb{S}\sim\pi/(\tilde{\lambda}T_{2}^{3/2})$ under the quantum control and
$\mathbb{S}_{\mathrm{free}}\sim\omega/(\tilde{\lambda}\sqrt{T_{2}})$ without
any control. Therefore, the adaptive quantum control enhances the sensitivity
by a factor
\[
\frac{\mathbb{S}_{\mathrm{free}}}{\mathbb{S}}\sim\frac{\omega T_{2}}{\pi}.
\]
For electron spin qubits in diamond nitrogen-vacancy center, the coherence
time $T_{2}$ reaches a few milliseconds
\cite{GaebelNature2006,TakahashiPRL2008,BalasubramanianNatMater2009,LangeScience2010}
at room temperature and even approaches one second at 77 K
\cite{BarGillNatCommun2013}. The experimentally demonstrated oscillator
frequency $\omega$ ranges from kHz to GHz (see Ref.
\onlinecite{AspelmeyerRMP2014} for a review). For a rough estimate, we take
$\omega/2\pi=100$ MHz and $T_{2}=1$ ms, which gives an enhancement
$\mathbb{S}_{\mathrm{free}}/\mathbb{S}\sim10^{5}$.

In summary, based on an exactly solvable qubit-oscillator model, we have
demonstrated theoretically the possibility to qualitatively improve the time
scaling of the sensing precision for the oscillator \ frequency from $1/T$ to
$1/T^{2}$ by applying adaptive periodic quantum control on the qubit,
without\textit{ }any direct access (initialization, control, or measurement)
to the oscillator. This improvement is applicable to a general initial states
of the oscillator and does not require appreciable qubit-oscillator
entanglement or net energy injection into the final state of the oscillator.
This provides a paradigm in which adaptive, periodic quantum control and
quantum backaction are utilized to steer the evolution of the target quantum
object and improve the precision of realistic quantum sensing by several
orders of magnitudes. Our study highlights a useful routine for high-precision
quantum sensing of remote quantum objects weakly coupled to a single quantum probe.

\section{Methods}

Here we outline the adaptive scheme that lifts the limitation Eq.
(\ref{CONSTRAINT}). Further details can be found in the supplementary
materials. The entire scheme consists of two stages:\ stage (i) and stage (ii).

Stage (i) corresponds to the uncertainty $\delta\omega$ satisfying $\omega
_{0}\gg\delta\omega\gtrsim\tilde{\lambda}$. In this stage, the large
uncertainty $\delta\omega$ only allows short evolution time $T$, so a single
measurement only improves the precision slightly. In the first step, we set
the evolution time to $T_{1}\sim2\pi/\delta\omega_{0}$ and perform $\nu
_{1}=c^{2}/G_{1}^{2}$ $(c$ is a constant controlling parameter and $G_{1}%
\sim\tilde{\lambda}/\delta\omega_{0}\ll1$) repeated measurements to improve
the precision from $\delta\omega_{0}$ to $\delta\omega_{1}\approx\delta
\omega_{0}/\sqrt{1+c^{2}}$. In the second step, we increase the evolution time
to $T_{2}\sim2\pi/\delta\omega_{1}\approx\sqrt{1+c^{2}}T_{1}$ and perform
$\nu_{2}\approx\nu_{1}/(1+c^{2})$ repeated measurements to improve the
precision to $\delta\omega_{2}\approx\delta\omega_{1}/\sqrt{1+c^{2}}$, and so
on, until the precision $\delta\omega$ becomes comparable or less than
$\tilde{\lambda}$. We denote the final estimator of this state by
$\omega_{\mathrm{i}}$ and its uncertainty by $\delta\omega_{\mathrm{i}}$. For
$c\ll1$, the total time cost of this stage is $T_{\mathrm{i}}\sim\delta
\omega_{0}/\tilde{\lambda}^{2}$ for $c\ll1$.

Stage (ii) corresponds to the uncertainty $\delta\omega\lesssim\tilde{\lambda
},$ which allows long evolution time, so a single measurement can improve the
precision signfiicantly. In the first step, we set the evolution time to
$T_{1}=(1/\kappa)\sqrt{2\pi/(\tilde{\lambda}\delta\omega_{\mathrm{i}})}$,
where $\kappa\gg1$ is a control parameter. Then we perform $\nu$ repeated
measurements to improve the precision to $\delta\omega_{1}\approx\delta
\omega_{\mathrm{i}}/\sqrt{1+\nu\eta^{2}}$, where $\eta\approx2/\kappa^{2}$. In
the second step, we increase the evolution time to $T_{2}\approx(1+\nu\eta
^{2})^{1/4}T_{1}$ and perform $\nu$ repeated measurements to improve the
precision to $\delta\omega_{2}\approx\delta\omega_{1}/\sqrt{1+\nu\eta^{2}}$,
and so on. At the end of the $m$th step, the total time cost is
$T_{\mathrm{ii}}=\nu(T_{1}+\cdots+T_{m})$ and the final precision is
$\delta\omega\approx\delta\omega_{\mathrm{i}}/(\sqrt{1+\nu\eta^{2}})^{m}$. For
$\sqrt{\nu}\eta\ll1$, we have%
\[
\delta\omega\approx\frac{16\pi}{\eta^{3}}\frac{1}{\tilde{\lambda
}T_{\mathrm{ii}}^{2}}\sim\frac{1}{\tilde{\lambda}T_{\mathrm{ii}}^{2}}.
\]
The total time cost of both stages is $\mathbb{T}\equiv T_{\mathrm{i}%
}+T_{\mathrm{ii}}$. When $\delta\omega_{0}$ is not too large compared with
$\tilde{\lambda}$ and/or the desired final precision is high, we have
$\mathbb{T}\approx T_{\mathrm{ii}}$, so the sensing precision follows
$1/\mathbb{T}^{2}$ scaling with the total time cost.


\begin{thebibliography}{60}
\expandafter\ifx\csname natexlab\endcsname\relax\def\natexlab#1{#1}\fi
\expandafter\ifx\csname bibnamefont\endcsname\relax
  \def\bibnamefont#1{#1}\fi
\expandafter\ifx\csname bibfnamefont\endcsname\relax
  \def\bibfnamefont#1{#1}\fi
\expandafter\ifx\csname citenamefont\endcsname\relax
  \def\citenamefont#1{#1}\fi
\expandafter\ifx\csname url\endcsname\relax
  \def\url#1{\texttt{#1}}\fi
\expandafter\ifx\csname urlprefix\endcsname\relax\def\urlprefix{URL }\fi
\providecommand{\bibinfo}[2]{#2}
\providecommand{\eprint}[2][]{\url{#2}}

\bibitem[{\citenamefont{Taylor et~al.}(2008)\citenamefont{Taylor, Cappellaro,
  Childress, Jiang, Budker, Hemmer, Yacoby, Walsworth, and
  Lukin}}]{TaylorNatPhys2008}
\bibinfo{author}{\bibfnamefont{J.~M.} \bibnamefont{Taylor}},
  \bibinfo{author}{\bibfnamefont{P.}~\bibnamefont{Cappellaro}},
  \bibinfo{author}{\bibfnamefont{L.}~\bibnamefont{Childress}},
  \bibinfo{author}{\bibfnamefont{L.}~\bibnamefont{Jiang}},
  \bibinfo{author}{\bibfnamefont{D.}~\bibnamefont{Budker}},
  \bibinfo{author}{\bibfnamefont{P.~R.} \bibnamefont{Hemmer}},
  \bibinfo{author}{\bibfnamefont{A.}~\bibnamefont{Yacoby}},
  \bibinfo{author}{\bibfnamefont{R.}~\bibnamefont{Walsworth}},
  \bibnamefont{and} \bibinfo{author}{\bibfnamefont{M.~D.} \bibnamefont{Lukin}},
  \bibinfo{journal}{Nat. Phys.} \textbf{\bibinfo{volume}{4}},
  \bibinfo{pages}{810} (\bibinfo{year}{2008}).

\bibitem[{\citenamefont{Balasubramanian
  et~al.}(2008)\citenamefont{Balasubramanian, Chan, Kolesov, Al-Hmoud, Tisler,
  Shin, Kim, Wojcik, Hemmer, Krueger et~al.}}]{BalasubramanianNature2008}
\bibinfo{author}{\bibfnamefont{G.}~\bibnamefont{Balasubramanian}},
  \bibinfo{author}{\bibfnamefont{I.~Y.} \bibnamefont{Chan}},
  \bibinfo{author}{\bibfnamefont{R.}~\bibnamefont{Kolesov}},
  \bibinfo{author}{\bibfnamefont{M.}~\bibnamefont{Al-Hmoud}},
  \bibinfo{author}{\bibfnamefont{J.}~\bibnamefont{Tisler}},
  \bibinfo{author}{\bibfnamefont{C.}~\bibnamefont{Shin}},
  \bibinfo{author}{\bibfnamefont{C.}~\bibnamefont{Kim}},
  \bibinfo{author}{\bibfnamefont{A.}~\bibnamefont{Wojcik}},
  \bibinfo{author}{\bibfnamefont{P.~R.} \bibnamefont{Hemmer}},
  \bibinfo{author}{\bibfnamefont{A.}~\bibnamefont{Krueger}},
  \bibnamefont{et~al.}, \bibinfo{journal}{Nature}
  \textbf{\bibinfo{volume}{455}}, \bibinfo{pages}{648} (\bibinfo{year}{2008}).

\bibitem[{\citenamefont{Maze et~al.}(2008)\citenamefont{Maze, Stanwix, Hodges,
  Hong, Taylor, Cappellaro, Jiang, Dutt, Togan, Zibrov
  et~al.}}]{MazeNature2008}
\bibinfo{author}{\bibfnamefont{J.~R.} \bibnamefont{Maze}},
  \bibinfo{author}{\bibfnamefont{P.~L.} \bibnamefont{Stanwix}},
  \bibinfo{author}{\bibfnamefont{J.~S.} \bibnamefont{Hodges}},
  \bibinfo{author}{\bibfnamefont{S.}~\bibnamefont{Hong}},
  \bibinfo{author}{\bibfnamefont{J.~M.} \bibnamefont{Taylor}},
  \bibinfo{author}{\bibfnamefont{P.}~\bibnamefont{Cappellaro}},
  \bibinfo{author}{\bibfnamefont{L.}~\bibnamefont{Jiang}},
  \bibinfo{author}{\bibfnamefont{M.~V.~G.} \bibnamefont{Dutt}},
  \bibinfo{author}{\bibfnamefont{E.}~\bibnamefont{Togan}},
  \bibinfo{author}{\bibfnamefont{A.~S.} \bibnamefont{Zibrov}},
  \bibnamefont{et~al.}, \bibinfo{journal}{Nature}
  \textbf{\bibinfo{volume}{455}}, \bibinfo{pages}{644} (\bibinfo{year}{2008}).

\bibitem[{\citenamefont{Robledo et~al.}(2011)\citenamefont{Robledo, Childress,
  Bernien, Hensen, Alkemade, and Hanson}}]{RobledoNature2011}
\bibinfo{author}{\bibfnamefont{L.}~\bibnamefont{Robledo}},
  \bibinfo{author}{\bibfnamefont{L.}~\bibnamefont{Childress}},
  \bibinfo{author}{\bibfnamefont{H.}~\bibnamefont{Bernien}},
  \bibinfo{author}{\bibfnamefont{B.}~\bibnamefont{Hensen}},
  \bibinfo{author}{\bibfnamefont{P.~F.~A.} \bibnamefont{Alkemade}},
  \bibnamefont{and} \bibinfo{author}{\bibfnamefont{R.}~\bibnamefont{Hanson}},
  \bibinfo{journal}{Nature} \textbf{\bibinfo{volume}{477}},
  \bibinfo{pages}{574} (\bibinfo{year}{2011}).

\bibitem[{\citenamefont{Togan et~al.}(2010)\citenamefont{Togan, Chu, Trifonov,
  Jiang, Maze, Childress, Dutt, Sorensen, Hemmer, Zibrov
  et~al.}}]{ToganNature2010}
\bibinfo{author}{\bibfnamefont{E.}~\bibnamefont{Togan}},
  \bibinfo{author}{\bibfnamefont{Y.}~\bibnamefont{Chu}},
  \bibinfo{author}{\bibfnamefont{A.~S.} \bibnamefont{Trifonov}},
  \bibinfo{author}{\bibfnamefont{L.}~\bibnamefont{Jiang}},
  \bibinfo{author}{\bibfnamefont{J.}~\bibnamefont{Maze}},
  \bibinfo{author}{\bibfnamefont{L.}~\bibnamefont{Childress}},
  \bibinfo{author}{\bibfnamefont{M.~V.~G.} \bibnamefont{Dutt}},
  \bibinfo{author}{\bibfnamefont{A.~S.} \bibnamefont{Sorensen}},
  \bibinfo{author}{\bibfnamefont{P.~R.} \bibnamefont{Hemmer}},
  \bibinfo{author}{\bibfnamefont{A.~S.} \bibnamefont{Zibrov}},
  \bibnamefont{et~al.}, \bibinfo{journal}{Nature}
  \textbf{\bibinfo{volume}{466}}, \bibinfo{pages}{730} (\bibinfo{year}{2010}).

\bibitem[{\citenamefont{Togan et~al.}(2011)\citenamefont{Togan, Chu, Imamoglu,
  and Lukin}}]{ToganNature2011}
\bibinfo{author}{\bibfnamefont{E.}~\bibnamefont{Togan}},
  \bibinfo{author}{\bibfnamefont{Y.}~\bibnamefont{Chu}},
  \bibinfo{author}{\bibfnamefont{A.}~\bibnamefont{Imamoglu}}, \bibnamefont{and}
  \bibinfo{author}{\bibfnamefont{M.~D.} \bibnamefont{Lukin}},
  \bibinfo{journal}{Nature} \textbf{\bibinfo{volume}{478}},
  \bibinfo{pages}{497} (\bibinfo{year}{2011}).

\bibitem[{\citenamefont{Maze et~al.}(2011)\citenamefont{Maze, Gali, Togan, Chu,
  Trifonov, Kaxiras, and Lukin}}]{MazeNJP2011}
\bibinfo{author}{\bibfnamefont{J.~R.} \bibnamefont{Maze}},
  \bibinfo{author}{\bibfnamefont{A.}~\bibnamefont{Gali}},
  \bibinfo{author}{\bibfnamefont{E.}~\bibnamefont{Togan}},
  \bibinfo{author}{\bibfnamefont{Y.}~\bibnamefont{Chu}},
  \bibinfo{author}{\bibfnamefont{A.}~\bibnamefont{Trifonov}},
  \bibinfo{author}{\bibfnamefont{E.}~\bibnamefont{Kaxiras}}, \bibnamefont{and}
  \bibinfo{author}{\bibfnamefont{M.~D.} \bibnamefont{Lukin}},
  \bibinfo{journal}{New J. Phys.} \textbf{\bibinfo{volume}{13}},
  \bibinfo{pages}{025025} (\bibinfo{year}{2011}).

\bibitem[{\citenamefont{O'Brien et~al.}(2001)\citenamefont{O'Brien, Schofield,
  Simmons, Clark, Dzurak, Curson, Kane, McAlpine, Hawley, and
  Brown}}]{OBrienPRB2001}
\bibinfo{author}{\bibfnamefont{J.~L.} \bibnamefont{O'Brien}},
  \bibinfo{author}{\bibfnamefont{S.~R.} \bibnamefont{Schofield}},
  \bibinfo{author}{\bibfnamefont{M.~Y.} \bibnamefont{Simmons}},
  \bibinfo{author}{\bibfnamefont{R.~G.} \bibnamefont{Clark}},
  \bibinfo{author}{\bibfnamefont{A.~S.} \bibnamefont{Dzurak}},
  \bibinfo{author}{\bibfnamefont{N.~J.} \bibnamefont{Curson}},
  \bibinfo{author}{\bibfnamefont{B.~E.} \bibnamefont{Kane}},
  \bibinfo{author}{\bibfnamefont{N.~S.} \bibnamefont{McAlpine}},
  \bibinfo{author}{\bibfnamefont{M.~E.} \bibnamefont{Hawley}},
  \bibnamefont{and} \bibinfo{author}{\bibfnamefont{G.~W.} \bibnamefont{Brown}},
  \bibinfo{journal}{Phys. Rev. B} \textbf{\bibinfo{volume}{64}},
  \bibinfo{pages}{161401} (\bibinfo{year}{2001}).

\bibitem[{\citenamefont{Morello et~al.}(2010)\citenamefont{Morello, Pla,
  Zwanenburg, Chan, Tan, Huebl, Mottonen, Nugroho, Yang, van Donkelaar
  et~al.}}]{MorelloNature2010}
\bibinfo{author}{\bibfnamefont{A.}~\bibnamefont{Morello}},
  \bibinfo{author}{\bibfnamefont{J.~J.} \bibnamefont{Pla}},
  \bibinfo{author}{\bibfnamefont{F.~A.} \bibnamefont{Zwanenburg}},
  \bibinfo{author}{\bibfnamefont{K.~W.} \bibnamefont{Chan}},
  \bibinfo{author}{\bibfnamefont{K.~Y.} \bibnamefont{Tan}},
  \bibinfo{author}{\bibfnamefont{H.}~\bibnamefont{Huebl}},
  \bibinfo{author}{\bibfnamefont{M.}~\bibnamefont{Mottonen}},
  \bibinfo{author}{\bibfnamefont{C.~D.} \bibnamefont{Nugroho}},
  \bibinfo{author}{\bibfnamefont{C.}~\bibnamefont{Yang}},
  \bibinfo{author}{\bibfnamefont{J.~A.} \bibnamefont{van Donkelaar}},
  \bibnamefont{et~al.}, \bibinfo{journal}{Nature}
  \textbf{\bibinfo{volume}{467}}, \bibinfo{pages}{687} (\bibinfo{year}{2010}).

\bibitem[{\citenamefont{Degen et~al.}(2017)\citenamefont{Degen, Reinhard, and
  Cappellaro}}]{DegenRMP2017}
\bibinfo{author}{\bibfnamefont{C.~L.} \bibnamefont{Degen}},
  \bibinfo{author}{\bibfnamefont{F.}~\bibnamefont{Reinhard}}, \bibnamefont{and}
  \bibinfo{author}{\bibfnamefont{P.}~\bibnamefont{Cappellaro}},
  \bibinfo{journal}{Rev. Mod. Phys.} \textbf{\bibinfo{volume}{89}},
  \bibinfo{pages}{035002} (\bibinfo{year}{2017}).

\bibitem[{\citenamefont{Zhao et~al.}(2011)\citenamefont{Zhao, Hu, Ho, Wan, and
  Liu}}]{ZhaoNatNano2011}
\bibinfo{author}{\bibfnamefont{N.}~\bibnamefont{Zhao}},
  \bibinfo{author}{\bibfnamefont{J.-L.} \bibnamefont{Hu}},
  \bibinfo{author}{\bibfnamefont{S.-W.} \bibnamefont{Ho}},
  \bibinfo{author}{\bibfnamefont{J.~T.~K.} \bibnamefont{Wan}},
  \bibnamefont{and} \bibinfo{author}{\bibfnamefont{R.-B.} \bibnamefont{Liu}},
  \bibinfo{journal}{Nat. Nanotechnol.} \textbf{\bibinfo{volume}{6}},
  \bibinfo{pages}{242} (\bibinfo{year}{2011}).

\bibitem[{\citenamefont{Zhao et~al.}(2012)\citenamefont{Zhao, Honert, Schmid,
  Klas, Isoya, Markham, Twitchen, Jelezko, Liu, Fedder
  et~al.}}]{ZhaoNatNano2012}
\bibinfo{author}{\bibfnamefont{N.}~\bibnamefont{Zhao}},
  \bibinfo{author}{\bibfnamefont{J.}~\bibnamefont{Honert}},
  \bibinfo{author}{\bibfnamefont{B.}~\bibnamefont{Schmid}},
  \bibinfo{author}{\bibfnamefont{M.}~\bibnamefont{Klas}},
  \bibinfo{author}{\bibfnamefont{J.}~\bibnamefont{Isoya}},
  \bibinfo{author}{\bibfnamefont{M.}~\bibnamefont{Markham}},
  \bibinfo{author}{\bibfnamefont{D.}~\bibnamefont{Twitchen}},
  \bibinfo{author}{\bibfnamefont{F.}~\bibnamefont{Jelezko}},
  \bibinfo{author}{\bibfnamefont{R.-B.} \bibnamefont{Liu}},
  \bibinfo{author}{\bibfnamefont{H.}~\bibnamefont{Fedder}},
  \bibnamefont{et~al.}, \bibinfo{journal}{Nat Nano}
  \textbf{\bibinfo{volume}{7}}, \bibinfo{pages}{657} (\bibinfo{year}{2012}).

\bibitem[{\citenamefont{Taminiau et~al.}(2012)\citenamefont{Taminiau, Wagenaar,
  van~der Sar, Jelezko, Dobrovitski, and Hanson}}]{TaminiauPRL2012}
\bibinfo{author}{\bibfnamefont{T.~H.} \bibnamefont{Taminiau}},
  \bibinfo{author}{\bibfnamefont{J.~J.~T.} \bibnamefont{Wagenaar}},
  \bibinfo{author}{\bibfnamefont{T.}~\bibnamefont{van~der Sar}},
  \bibinfo{author}{\bibfnamefont{F.}~\bibnamefont{Jelezko}},
  \bibinfo{author}{\bibfnamefont{V.~V.} \bibnamefont{Dobrovitski}},
  \bibnamefont{and} \bibinfo{author}{\bibfnamefont{R.}~\bibnamefont{Hanson}},
  \bibinfo{journal}{Phys. Rev. Lett.} \textbf{\bibinfo{volume}{109}},
  \bibinfo{pages}{137602} (\bibinfo{year}{2012}).

\bibitem[{\citenamefont{Kolkowitz
  et~al.}(2012{\natexlab{a}})\citenamefont{Kolkowitz, Unterreithmeier, Bennett,
  and Lukin}}]{KolkowitzPRL2012}
\bibinfo{author}{\bibfnamefont{S.}~\bibnamefont{Kolkowitz}},
  \bibinfo{author}{\bibfnamefont{Q.~P.} \bibnamefont{Unterreithmeier}},
  \bibinfo{author}{\bibfnamefont{S.~D.} \bibnamefont{Bennett}},
  \bibnamefont{and} \bibinfo{author}{\bibfnamefont{M.~D.} \bibnamefont{Lukin}},
  \bibinfo{journal}{Phys. Rev. Lett.} \textbf{\bibinfo{volume}{109}},
  \bibinfo{pages}{137601} (\bibinfo{year}{2012}{\natexlab{a}}).

\bibitem[{\citenamefont{Cai et~al.}(2013)\citenamefont{Cai, Jelezko, Plenio,
  and Retzker}}]{CaiNJP2013}
\bibinfo{author}{\bibfnamefont{J.}~\bibnamefont{Cai}},
  \bibinfo{author}{\bibfnamefont{F.}~\bibnamefont{Jelezko}},
  \bibinfo{author}{\bibfnamefont{M.~B.} \bibnamefont{Plenio}},
  \bibnamefont{and} \bibinfo{author}{\bibfnamefont{A.}~\bibnamefont{Retzker}},
  \bibinfo{journal}{New J. Phys.} \textbf{\bibinfo{volume}{15}},
  \bibinfo{pages}{013020} (\bibinfo{year}{2013}).

\bibitem[{\citenamefont{London et~al.}(2013)\citenamefont{London, Scheuer, Cai,
  Schwarz, Retzker, Plenio, Katagiri, Teraji, Koizumi, Isoya
  et~al.}}]{LondonPRL2013}
\bibinfo{author}{\bibfnamefont{P.}~\bibnamefont{London}},
  \bibinfo{author}{\bibfnamefont{J.}~\bibnamefont{Scheuer}},
  \bibinfo{author}{\bibfnamefont{J.-M.} \bibnamefont{Cai}},
  \bibinfo{author}{\bibfnamefont{I.}~\bibnamefont{Schwarz}},
  \bibinfo{author}{\bibfnamefont{A.}~\bibnamefont{Retzker}},
  \bibinfo{author}{\bibfnamefont{M.~B.} \bibnamefont{Plenio}},
  \bibinfo{author}{\bibfnamefont{M.}~\bibnamefont{Katagiri}},
  \bibinfo{author}{\bibfnamefont{T.}~\bibnamefont{Teraji}},
  \bibinfo{author}{\bibfnamefont{S.}~\bibnamefont{Koizumi}},
  \bibinfo{author}{\bibfnamefont{J.}~\bibnamefont{Isoya}},
  \bibnamefont{et~al.}, \bibinfo{journal}{Phys. Rev. Lett.}
  \textbf{\bibinfo{volume}{111}}, \bibinfo{pages}{067601}
  (\bibinfo{year}{2013}).

\bibitem[{\citenamefont{Laraoui et~al.}(2013)\citenamefont{Laraoui, Dolde,
  Burk, Reinhard, Wrachtrup, and Meriles}}]{LaraouiNatCommun2013}
\bibinfo{author}{\bibfnamefont{A.}~\bibnamefont{Laraoui}},
  \bibinfo{author}{\bibfnamefont{F.}~\bibnamefont{Dolde}},
  \bibinfo{author}{\bibfnamefont{C.}~\bibnamefont{Burk}},
  \bibinfo{author}{\bibfnamefont{F.}~\bibnamefont{Reinhard}},
  \bibinfo{author}{\bibfnamefont{J.}~\bibnamefont{Wrachtrup}},
  \bibnamefont{and} \bibinfo{author}{\bibfnamefont{C.~A.}
  \bibnamefont{Meriles}}, \bibinfo{journal}{Nat. Commun.}
  \textbf{\bibinfo{volume}{4}}, \bibinfo{pages}{1651} (\bibinfo{year}{2013}).

\bibitem[{\citenamefont{Shi et~al.}(2014)\citenamefont{Shi, Kong, Wang, Kong,
  Zhao, Liu, and Du}}]{ShiNatPhys2014}
\bibinfo{author}{\bibfnamefont{F.}~\bibnamefont{Shi}},
  \bibinfo{author}{\bibfnamefont{X.}~\bibnamefont{Kong}},
  \bibinfo{author}{\bibfnamefont{P.}~\bibnamefont{Wang}},
  \bibinfo{author}{\bibfnamefont{F.}~\bibnamefont{Kong}},
  \bibinfo{author}{\bibfnamefont{N.}~\bibnamefont{Zhao}},
  \bibinfo{author}{\bibfnamefont{R.-B.} \bibnamefont{Liu}}, \bibnamefont{and}
  \bibinfo{author}{\bibfnamefont{J.}~\bibnamefont{Du}}, \bibinfo{journal}{Nat.
  Phys.} \textbf{\bibinfo{volume}{10}}, \bibinfo{pages}{21}
  (\bibinfo{year}{2014}).

\bibitem[{\citenamefont{Lang et~al.}(2015)\citenamefont{Lang, Liu, and
  Monteiro}}]{LangPRX2015}
\bibinfo{author}{\bibfnamefont{J.~E.} \bibnamefont{Lang}},
  \bibinfo{author}{\bibfnamefont{R.~B.} \bibnamefont{Liu}}, \bibnamefont{and}
  \bibinfo{author}{\bibfnamefont{T.~S.} \bibnamefont{Monteiro}},
  \bibinfo{journal}{Phys. Rev. X} \textbf{\bibinfo{volume}{5}},
  \bibinfo{pages}{041016} (\bibinfo{year}{2015}).

\bibitem[{\citenamefont{Boss et~al.}(2016)\citenamefont{Boss, Chang, Armijo,
  Cujia, Rosskopf, Maze, and Degen}}]{BossPRL2016}
\bibinfo{author}{\bibfnamefont{J.~M.} \bibnamefont{Boss}},
  \bibinfo{author}{\bibfnamefont{K.}~\bibnamefont{Chang}},
  \bibinfo{author}{\bibfnamefont{J.}~\bibnamefont{Armijo}},
  \bibinfo{author}{\bibfnamefont{K.}~\bibnamefont{Cujia}},
  \bibinfo{author}{\bibfnamefont{T.}~\bibnamefont{Rosskopf}},
  \bibinfo{author}{\bibfnamefont{J.~R.} \bibnamefont{Maze}}, \bibnamefont{and}
  \bibinfo{author}{\bibfnamefont{C.~L.} \bibnamefont{Degen}},
  \bibinfo{journal}{Phys. Rev. Lett.} \textbf{\bibinfo{volume}{116}},
  \bibinfo{pages}{197601} (\bibinfo{year}{2016}).

\bibitem[{\citenamefont{Zaiser et~al.}(2016)\citenamefont{Zaiser, Rendler,
  Jakobi, Wolf, Lee, Wagner, Bergholm, Schulte-Herbrüggen, Neumann, and
  Wrachtrup}}]{ZaiserNC2016}
\bibinfo{author}{\bibfnamefont{S.}~\bibnamefont{Zaiser}},
  \bibinfo{author}{\bibfnamefont{T.}~\bibnamefont{Rendler}},
  \bibinfo{author}{\bibfnamefont{I.}~\bibnamefont{Jakobi}},
  \bibinfo{author}{\bibfnamefont{T.}~\bibnamefont{Wolf}},
  \bibinfo{author}{\bibfnamefont{S.-Y.} \bibnamefont{Lee}},
  \bibinfo{author}{\bibfnamefont{S.}~\bibnamefont{Wagner}},
  \bibinfo{author}{\bibfnamefont{V.}~\bibnamefont{Bergholm}},
  \bibinfo{author}{\bibfnamefont{T.}~\bibnamefont{Schulte-Herbrüggen}},
  \bibinfo{author}{\bibfnamefont{P.}~\bibnamefont{Neumann}}, \bibnamefont{and}
  \bibinfo{author}{\bibfnamefont{J.}~\bibnamefont{Wrachtrup}},
  \bibinfo{journal}{Nat. Commun.} \textbf{\bibinfo{volume}{7}},
  \bibinfo{pages}{12279} (\bibinfo{year}{2016}).

\bibitem[{\citenamefont{Ma and Liu}(2016{\natexlab{a}})}]{MaPRA2016}
\bibinfo{author}{\bibfnamefont{W.-L.} \bibnamefont{Ma}} \bibnamefont{and}
  \bibinfo{author}{\bibfnamefont{R.-B.} \bibnamefont{Liu}},
  \bibinfo{journal}{Phys. Rev. Applied} \textbf{\bibinfo{volume}{6}},
  \bibinfo{pages}{054012} (\bibinfo{year}{2016}{\natexlab{a}}).

\bibitem[{\citenamefont{Ma and Liu}(2016{\natexlab{b}})}]{MaPRA2016a}
\bibinfo{author}{\bibfnamefont{W.-L.} \bibnamefont{Ma}} \bibnamefont{and}
  \bibinfo{author}{\bibfnamefont{R.-B.} \bibnamefont{Liu}},
  \bibinfo{journal}{Phys. Rev. Applied} \textbf{\bibinfo{volume}{6}},
  \bibinfo{pages}{024019} (\bibinfo{year}{2016}{\natexlab{b}}).

\bibitem[{\citenamefont{Shu et~al.}(2017)\citenamefont{Shu, Zhang, Cao, Yang,
  Plenio, M\"uller, Lang, Tomek, Naydenov, McGuinness et~al.}}]{ShuPRA2017}
\bibinfo{author}{\bibfnamefont{Z.}~\bibnamefont{Shu}},
  \bibinfo{author}{\bibfnamefont{Z.}~\bibnamefont{Zhang}},
  \bibinfo{author}{\bibfnamefont{Q.}~\bibnamefont{Cao}},
  \bibinfo{author}{\bibfnamefont{P.}~\bibnamefont{Yang}},
  \bibinfo{author}{\bibfnamefont{M.~B.} \bibnamefont{Plenio}},
  \bibinfo{author}{\bibfnamefont{C.}~\bibnamefont{M\"uller}},
  \bibinfo{author}{\bibfnamefont{J.}~\bibnamefont{Lang}},
  \bibinfo{author}{\bibfnamefont{N.}~\bibnamefont{Tomek}},
  \bibinfo{author}{\bibfnamefont{B.}~\bibnamefont{Naydenov}},
  \bibinfo{author}{\bibfnamefont{L.~P.} \bibnamefont{McGuinness}},
  \bibnamefont{et~al.}, \bibinfo{journal}{Phys. Rev. A}
  \textbf{\bibinfo{volume}{96}}, \bibinfo{pages}{051402}
  (\bibinfo{year}{2017}).

\bibitem[{\citenamefont{Liu et~al.}(2017)\citenamefont{Liu, Plenio, and
  Cai}}]{LiuPRL2017a}
\bibinfo{author}{\bibfnamefont{H.}~\bibnamefont{Liu}},
  \bibinfo{author}{\bibfnamefont{M.~B.} \bibnamefont{Plenio}},
  \bibnamefont{and} \bibinfo{author}{\bibfnamefont{J.}~\bibnamefont{Cai}},
  \bibinfo{journal}{Phys. Rev. Lett.} \textbf{\bibinfo{volume}{118}},
  \bibinfo{pages}{200402} (\bibinfo{year}{2017}).

\bibitem[{\citenamefont{Yuan and Fung}(2015)}]{YuanPRL2015}
\bibinfo{author}{\bibfnamefont{H.}~\bibnamefont{Yuan}} \bibnamefont{and}
  \bibinfo{author}{\bibfnamefont{C.-H.~F.} \bibnamefont{Fung}},
  \bibinfo{journal}{Phys. Rev. Lett.} \textbf{\bibinfo{volume}{115}},
  \bibinfo{pages}{110401} (\bibinfo{year}{2015}).

\bibitem[{\citenamefont{Yuan}(2016)}]{YuanPRL2016}
\bibinfo{author}{\bibfnamefont{H.}~\bibnamefont{Yuan}}, \bibinfo{journal}{Phys.
  Rev. Lett.} \textbf{\bibinfo{volume}{117}}, \bibinfo{pages}{160801}
  (\bibinfo{year}{2016}).

\bibitem[{\citenamefont{Pang and Jordan}(2017)}]{PangNC2017}
\bibinfo{author}{\bibfnamefont{S.}~\bibnamefont{Pang}} \bibnamefont{and}
  \bibinfo{author}{\bibfnamefont{A.~N.} \bibnamefont{Jordan}},
  \bibinfo{journal}{Nat. Commun.} \textbf{\bibinfo{volume}{8}},
  \bibinfo{pages}{14695} (\bibinfo{year}{2017}).

\bibitem[{\citenamefont{Nusran et~al.}(2012)\citenamefont{Nusran, Momeen, , and
  Dutt}}]{NusranNatNano2012}
\bibinfo{author}{\bibfnamefont{N.~M.} \bibnamefont{Nusran}},
  \bibinfo{author}{\bibfnamefont{M.~U.} \bibnamefont{Momeen}}, ,
  \bibnamefont{and} \bibinfo{author}{\bibfnamefont{M.~V.~G.}
  \bibnamefont{Dutt}}, \bibinfo{journal}{Nat. Nanotechnol.}
  \textbf{\bibinfo{volume}{7}}, \bibinfo{pages}{109} (\bibinfo{year}{2012}).

\bibitem[{\citenamefont{Waldherr et~al.}(2012)\citenamefont{Waldherr, Beck,
  Neumann, Said, Nitsche, Markham, Twitchen, Twamley, Jelezko, and
  Wrachtrup}}]{WaldherrNatNano2012}
\bibinfo{author}{\bibfnamefont{G.}~\bibnamefont{Waldherr}},
  \bibinfo{author}{\bibfnamefont{J.}~\bibnamefont{Beck}},
  \bibinfo{author}{\bibfnamefont{P.}~\bibnamefont{Neumann}},
  \bibinfo{author}{\bibfnamefont{R.~S.} \bibnamefont{Said}},
  \bibinfo{author}{\bibfnamefont{M.}~\bibnamefont{Nitsche}},
  \bibinfo{author}{\bibfnamefont{M.~L.} \bibnamefont{Markham}},
  \bibinfo{author}{\bibfnamefont{D.~J.} \bibnamefont{Twitchen}},
  \bibinfo{author}{\bibfnamefont{J.}~\bibnamefont{Twamley}},
  \bibinfo{author}{\bibfnamefont{F.}~\bibnamefont{Jelezko}}, \bibnamefont{and}
  \bibinfo{author}{\bibfnamefont{J.}~\bibnamefont{Wrachtrup}},
  \bibinfo{journal}{Nat. Nanotechnol.} \textbf{\bibinfo{volume}{7}},
  \bibinfo{pages}{105} (\bibinfo{year}{2012}).

\bibitem[{\citenamefont{Bonato et~al.}(2016)\citenamefont{Bonato, Blok, Dinani,
  Berry, Markham, Twitchen, and Hanson}}]{BonatoNatNano2016}
\bibinfo{author}{\bibfnamefont{C.}~\bibnamefont{Bonato}},
  \bibinfo{author}{\bibfnamefont{M.~S.} \bibnamefont{Blok}},
  \bibinfo{author}{\bibfnamefont{H.~T.} \bibnamefont{Dinani}},
  \bibinfo{author}{\bibfnamefont{D.~W.} \bibnamefont{Berry}},
  \bibinfo{author}{\bibfnamefont{M.~L.} \bibnamefont{Markham}},
  \bibinfo{author}{\bibfnamefont{D.~J.} \bibnamefont{Twitchen}},
  \bibnamefont{and} \bibinfo{author}{\bibfnamefont{R.}~\bibnamefont{Hanson}},
  \bibinfo{journal}{Nat. Nanotechnol.} \textbf{\bibinfo{volume}{11}},
  \bibinfo{pages}{247} (\bibinfo{year}{2016}).

\bibitem[{\citenamefont{Schmitt et~al.}(2017)\citenamefont{Schmitt, Gefen,
  St{\"u}rner, Unden, Wolff, M{\"u}ller, Scheuer, Naydenov, Markham, Pezzagna
  et~al.}}]{SchmittScience2017}
\bibinfo{author}{\bibfnamefont{S.}~\bibnamefont{Schmitt}},
  \bibinfo{author}{\bibfnamefont{T.}~\bibnamefont{Gefen}},
  \bibinfo{author}{\bibfnamefont{F.~M.} \bibnamefont{St{\"u}rner}},
  \bibinfo{author}{\bibfnamefont{T.}~\bibnamefont{Unden}},
  \bibinfo{author}{\bibfnamefont{G.}~\bibnamefont{Wolff}},
  \bibinfo{author}{\bibfnamefont{C.}~\bibnamefont{M{\"u}ller}},
  \bibinfo{author}{\bibfnamefont{J.}~\bibnamefont{Scheuer}},
  \bibinfo{author}{\bibfnamefont{B.}~\bibnamefont{Naydenov}},
  \bibinfo{author}{\bibfnamefont{M.}~\bibnamefont{Markham}},
  \bibinfo{author}{\bibfnamefont{S.}~\bibnamefont{Pezzagna}},
  \bibnamefont{et~al.}, \bibinfo{journal}{Science}
  \textbf{\bibinfo{volume}{356}}, \bibinfo{pages}{832} (\bibinfo{year}{2017}).

\bibitem[{\citenamefont{Boss et~al.}(2017)\citenamefont{Boss, Cujia, Zopes, and
  Degen}}]{BossScience2017}
\bibinfo{author}{\bibfnamefont{J.~M.} \bibnamefont{Boss}},
  \bibinfo{author}{\bibfnamefont{K.~S.} \bibnamefont{Cujia}},
  \bibinfo{author}{\bibfnamefont{J.}~\bibnamefont{Zopes}}, \bibnamefont{and}
  \bibinfo{author}{\bibfnamefont{C.~L.} \bibnamefont{Degen}},
  \bibinfo{journal}{Science} \textbf{\bibinfo{volume}{356}},
  \bibinfo{pages}{837} (\bibinfo{year}{2017}).

\bibitem[{\citenamefont{Aspelmeyer et~al.}(2014)\citenamefont{Aspelmeyer,
  Kippenberg, and Marquardt}}]{AspelmeyerRMP2014}
\bibinfo{author}{\bibfnamefont{M.}~\bibnamefont{Aspelmeyer}},
  \bibinfo{author}{\bibfnamefont{T.~J.} \bibnamefont{Kippenberg}},
  \bibnamefont{and}
  \bibinfo{author}{\bibfnamefont{F.}~\bibnamefont{Marquardt}},
  \bibinfo{journal}{Rev. Mod. Phys.} \textbf{\bibinfo{volume}{86}},
  \bibinfo{pages}{1391} (\bibinfo{year}{2014}).

\bibitem[{\citenamefont{Braunstein and Caves}(1994)}]{BraunsteinPRL1994}
\bibinfo{author}{\bibfnamefont{S.~L.} \bibnamefont{Braunstein}}
  \bibnamefont{and} \bibinfo{author}{\bibfnamefont{C.~M.} \bibnamefont{Caves}},
  \bibinfo{journal}{Phys. Rev. Lett.} \textbf{\bibinfo{volume}{72}},
  \bibinfo{pages}{3439} (\bibinfo{year}{1994}).

\bibitem[{\citenamefont{Kay}(1993)}]{KayBook1993}
\bibinfo{author}{\bibfnamefont{S.~M.} \bibnamefont{Kay}},
  \emph{\bibinfo{title}{Fundamentals of Statistical Signal Processing:
  Estimation Theory}} (\bibinfo{publisher}{Prentice-Hall},
  \bibinfo{year}{1993}).

\bibitem[{\citenamefont{Helstrom}(1976)}]{HelstromBook1976}
\bibinfo{author}{\bibfnamefont{C.~W.} \bibnamefont{Helstrom}},
  \emph{\bibinfo{title}{Quantum Detection and Estimation Theory}}
  (\bibinfo{publisher}{Academic press, New York}, \bibinfo{year}{1976}).

\bibitem[{\citenamefont{Giovannetti et~al.}(2006)\citenamefont{Giovannetti,
  Lloyd, and Maccone}}]{GiovannettiPRL2006}
\bibinfo{author}{\bibfnamefont{V.}~\bibnamefont{Giovannetti}},
  \bibinfo{author}{\bibfnamefont{S.}~\bibnamefont{Lloyd}}, \bibnamefont{and}
  \bibinfo{author}{\bibfnamefont{L.}~\bibnamefont{Maccone}},
  \bibinfo{journal}{Phys. Rev. Lett.} \textbf{\bibinfo{volume}{96}},
  \bibinfo{pages}{010401} (\bibinfo{year}{2006}).

\bibitem[{\citenamefont{Pang and Brun}(2014)}]{PangPRA2014}
\bibinfo{author}{\bibfnamefont{S.}~\bibnamefont{Pang}} \bibnamefont{and}
  \bibinfo{author}{\bibfnamefont{T.~A.} \bibnamefont{Brun}},
  \bibinfo{journal}{Phys. Rev. A} \textbf{\bibinfo{volume}{90}},
  \bibinfo{pages}{022117} (\bibinfo{year}{2014}).

\bibitem[{\citenamefont{Liu et~al.}(2015)\citenamefont{Liu, Jing, and
  Wang}}]{LiuSciRep2015}
\bibinfo{author}{\bibfnamefont{J.}~\bibnamefont{Liu}},
  \bibinfo{author}{\bibfnamefont{X.-X.} \bibnamefont{Jing}}, \bibnamefont{and}
  \bibinfo{author}{\bibfnamefont{X.}~\bibnamefont{Wang}},
  \bibinfo{journal}{Sci. Rep.} \textbf{\bibinfo{volume}{5}},
  \bibinfo{pages}{8565} (\bibinfo{year}{2015}).

\bibitem[{\citenamefont{Neukirch et~al.}(2013)\citenamefont{Neukirch, Gieseler,
  Quidant, Novotny, and Vamivakas}}]{NeukirchOL2013}
\bibinfo{author}{\bibfnamefont{L.~P.} \bibnamefont{Neukirch}},
  \bibinfo{author}{\bibfnamefont{J.}~\bibnamefont{Gieseler}},
  \bibinfo{author}{\bibfnamefont{R.}~\bibnamefont{Quidant}},
  \bibinfo{author}{\bibfnamefont{L.}~\bibnamefont{Novotny}}, \bibnamefont{and}
  \bibinfo{author}{\bibfnamefont{A.~N.} \bibnamefont{Vamivakas}},
  \bibinfo{journal}{Opt. Lett.} \textbf{\bibinfo{volume}{38}},
  \bibinfo{pages}{2976} (\bibinfo{year}{2013}).

\bibitem[{\citenamefont{Scala et~al.}(2013)\citenamefont{Scala, Kim, Morley,
  Barker, and Bose}}]{ScalaPRL2013}
\bibinfo{author}{\bibfnamefont{M.}~\bibnamefont{Scala}},
  \bibinfo{author}{\bibfnamefont{M.~S.} \bibnamefont{Kim}},
  \bibinfo{author}{\bibfnamefont{G.~W.} \bibnamefont{Morley}},
  \bibinfo{author}{\bibfnamefont{P.~F.} \bibnamefont{Barker}},
  \bibnamefont{and} \bibinfo{author}{\bibfnamefont{S.}~\bibnamefont{Bose}},
  \bibinfo{journal}{Phys. Rev. Lett.} \textbf{\bibinfo{volume}{111}},
  \bibinfo{pages}{180403} (\bibinfo{year}{2013}).

\bibitem[{\citenamefont{Zhao and Yin}(2014)}]{ZhaoPRA2014}
\bibinfo{author}{\bibfnamefont{N.}~\bibnamefont{Zhao}} \bibnamefont{and}
  \bibinfo{author}{\bibfnamefont{Z.~Q.} \bibnamefont{Yin}},
  \bibinfo{journal}{Phys. Rev. A} \textbf{\bibinfo{volume}{90}},
  \bibinfo{pages}{042118} (\bibinfo{year}{2014}).

\bibitem[{\citenamefont{LaHaye et~al.}(2009)\citenamefont{LaHaye, Suh,
  Echternach, Schwab, and Roukes}}]{LaHayeNature2009}
\bibinfo{author}{\bibfnamefont{M.~D.} \bibnamefont{LaHaye}},
  \bibinfo{author}{\bibfnamefont{J.}~\bibnamefont{Suh}},
  \bibinfo{author}{\bibfnamefont{P.~M.} \bibnamefont{Echternach}},
  \bibinfo{author}{\bibfnamefont{K.~C.} \bibnamefont{Schwab}},
  \bibnamefont{and} \bibinfo{author}{\bibfnamefont{M.~L.}
  \bibnamefont{Roukes}}, \bibinfo{journal}{Nature}
  \textbf{\bibinfo{volume}{459}}, \bibinfo{pages}{960} (\bibinfo{year}{2009}).

\bibitem[{\citenamefont{Hunger et~al.}(2010)\citenamefont{Hunger, Camerer,
  H\"ansch, K\"onig, Kotthaus, Reichel, and Treutlein}}]{HungerPRL2010}
\bibinfo{author}{\bibfnamefont{D.}~\bibnamefont{Hunger}},
  \bibinfo{author}{\bibfnamefont{S.}~\bibnamefont{Camerer}},
  \bibinfo{author}{\bibfnamefont{T.~W.} \bibnamefont{H\"ansch}},
  \bibinfo{author}{\bibfnamefont{D.}~\bibnamefont{K\"onig}},
  \bibinfo{author}{\bibfnamefont{J.~P.} \bibnamefont{Kotthaus}},
  \bibinfo{author}{\bibfnamefont{J.}~\bibnamefont{Reichel}}, \bibnamefont{and}
  \bibinfo{author}{\bibfnamefont{P.}~\bibnamefont{Treutlein}},
  \bibinfo{journal}{Phys. Rev. Lett.} \textbf{\bibinfo{volume}{104}},
  \bibinfo{pages}{143002} (\bibinfo{year}{2010}).

\bibitem[{\citenamefont{Bennett et~al.}(2010)\citenamefont{Bennett, Cockins,
  Miyahara, Gr\"utter, and Clerk}}]{BennettPRL2010}
\bibinfo{author}{\bibfnamefont{S.~D.} \bibnamefont{Bennett}},
  \bibinfo{author}{\bibfnamefont{L.}~\bibnamefont{Cockins}},
  \bibinfo{author}{\bibfnamefont{Y.}~\bibnamefont{Miyahara}},
  \bibinfo{author}{\bibfnamefont{P.}~\bibnamefont{Gr\"utter}},
  \bibnamefont{and} \bibinfo{author}{\bibfnamefont{A.~A.} \bibnamefont{Clerk}},
  \bibinfo{journal}{Phys. Rev. Lett.} \textbf{\bibinfo{volume}{104}},
  \bibinfo{pages}{017203} (\bibinfo{year}{2010}).

\bibitem[{\citenamefont{Arcizet et~al.}(2011)\citenamefont{Arcizet, Jacques,
  Siria, Poncharal, Vincent, and Seidelin}}]{ArcizetNatPhys2011}
\bibinfo{author}{\bibfnamefont{O.}~\bibnamefont{Arcizet}},
  \bibinfo{author}{\bibfnamefont{V.}~\bibnamefont{Jacques}},
  \bibinfo{author}{\bibfnamefont{A.}~\bibnamefont{Siria}},
  \bibinfo{author}{\bibfnamefont{P.}~\bibnamefont{Poncharal}},
  \bibinfo{author}{\bibfnamefont{P.}~\bibnamefont{Vincent}}, \bibnamefont{and}
  \bibinfo{author}{\bibfnamefont{S.}~\bibnamefont{Seidelin}},
  \bibinfo{journal}{Nat. Phys.} \textbf{\bibinfo{volume}{7}},
  \bibinfo{pages}{1} (\bibinfo{year}{2011}).

\bibitem[{\citenamefont{Kolkowitz
  et~al.}(2012{\natexlab{b}})\citenamefont{Kolkowitz, {Bleszynski Jayich},
  Unterreithmeier, Bennett, Rabl, Harris, and Lukin}}]{KolkowitzS2012}
\bibinfo{author}{\bibfnamefont{S.}~\bibnamefont{Kolkowitz}},
  \bibinfo{author}{\bibfnamefont{a.~C.} \bibnamefont{{Bleszynski Jayich}}},
  \bibinfo{author}{\bibfnamefont{Q.}~\bibnamefont{Unterreithmeier}},
  \bibinfo{author}{\bibfnamefont{S.~D.} \bibnamefont{Bennett}},
  \bibinfo{author}{\bibfnamefont{P.}~\bibnamefont{Rabl}},
  \bibinfo{author}{\bibfnamefont{J.~G.~E.} \bibnamefont{Harris}},
  \bibnamefont{and} \bibinfo{author}{\bibfnamefont{M.~D.} \bibnamefont{Lukin}},
  \bibinfo{journal}{Science} \textbf{\bibinfo{volume}{335}},
  \bibinfo{pages}{1603} (\bibinfo{year}{2012}{\natexlab{b}}).

\bibitem[{\citenamefont{YeoI. et~al.}(2014)\citenamefont{YeoI., de~AssisP-L.,
  GloppeA., Dupont-FerrierE., VerlotP., S., DupuyE., ClaudonJ., GerardJ-M.,
  AuffevesA. et~al.}}]{YeoINatNano2014}
\bibinfo{author}{\bibnamefont{YeoI.}},
  \bibinfo{author}{\bibnamefont{de~AssisP-L.}},
  \bibinfo{author}{\bibnamefont{GloppeA.}},
  \bibinfo{author}{\bibnamefont{Dupont-FerrierE.}},
  \bibinfo{author}{\bibnamefont{VerlotP.}},
  \bibinfo{author}{\bibfnamefont{M.}~\bibnamefont{S.}},
  \bibinfo{author}{\bibnamefont{DupuyE.}},
  \bibinfo{author}{\bibnamefont{ClaudonJ.}},
  \bibinfo{author}{\bibnamefont{GerardJ-M.}},
  \bibinfo{author}{\bibnamefont{AuffevesA.}}, \bibnamefont{et~al.},
  \bibinfo{journal}{Nat Nano} \textbf{\bibinfo{volume}{9}},
  \bibinfo{pages}{106} (\bibinfo{year}{2014}).

\bibitem[{\citenamefont{Yang et~al.}(2017)\citenamefont{Yang, Ma, and
  Liu}}]{YangRPP2017}
\bibinfo{author}{\bibfnamefont{W.}~\bibnamefont{Yang}},
  \bibinfo{author}{\bibfnamefont{W.-L.} \bibnamefont{Ma}}, \bibnamefont{and}
  \bibinfo{author}{\bibfnamefont{R.-B.} \bibnamefont{Liu}},
  \bibinfo{journal}{Rep. Prog. Phys.} \textbf{\bibinfo{volume}{80}},
  \bibinfo{pages}{016001} (\bibinfo{year}{2017}).

\bibitem[{\citenamefont{Sun and Xiao}(1991)}]{SunCTP1991}
\bibinfo{author}{\bibfnamefont{C.-P.} \bibnamefont{Sun}} \bibnamefont{and}
  \bibinfo{author}{\bibfnamefont{Q.}~\bibnamefont{Xiao}},
  \bibinfo{journal}{Communications in Theoretical Physics}
  \textbf{\bibinfo{volume}{16}}, \bibinfo{pages}{359} (\bibinfo{year}{1991}).

\bibitem[{\citenamefont{Zhong et~al.}(2013)\citenamefont{Zhong, Sun, Ma, Wang,
  and Nori}}]{ZhongPRA2013}
\bibinfo{author}{\bibfnamefont{W.}~\bibnamefont{Zhong}},
  \bibinfo{author}{\bibfnamefont{Z.}~\bibnamefont{Sun}},
  \bibinfo{author}{\bibfnamefont{J.}~\bibnamefont{Ma}},
  \bibinfo{author}{\bibfnamefont{X.}~\bibnamefont{Wang}}, \bibnamefont{and}
  \bibinfo{author}{\bibfnamefont{F.}~\bibnamefont{Nori}},
  \bibinfo{journal}{Phys. Rev. A} \textbf{\bibinfo{volume}{87}},
  \bibinfo{pages}{022337} (\bibinfo{year}{2013}).

\bibitem[{\citenamefont{Carr and Purcell}(1954)}]{CarrPR1954}
\bibinfo{author}{\bibfnamefont{H.}~\bibnamefont{Carr}} \bibnamefont{and}
  \bibinfo{author}{\bibfnamefont{E.~M.} \bibnamefont{Purcell}},
  \bibinfo{journal}{Phys. Rev.} \textbf{\bibinfo{volume}{94}},
  \bibinfo{pages}{630} (\bibinfo{year}{1954}).

\bibitem[{\citenamefont{Meiboom and Gill}(1958)}]{MeiboomRSI1958}
\bibinfo{author}{\bibfnamefont{S.}~\bibnamefont{Meiboom}} \bibnamefont{and}
  \bibinfo{author}{\bibfnamefont{D.}~\bibnamefont{Gill}},
  \bibinfo{journal}{Rev. Sci. Instrum.} \textbf{\bibinfo{volume}{29}},
  \bibinfo{pages}{688} (\bibinfo{year}{1958}).

\bibitem[{Note1()}]{Note1}
Note1, \bibinfo{note}{here we consider using the quantum probe for
  high-precision sensing of a well-defined frequency of the target quantum
  object. This requires the coherence time $T_{\protect \mathrm {tar}}$ of the
  target to be much longer than that of the quantum probe, otherwise the target
  frequency would be broadened by $\sim 1/T_{\protect \mathrm {tar}}$, making
  high-precision sensing impossible.}

\bibitem[{\citenamefont{Gaebel et~al.}(2006)\citenamefont{Gaebel, Domhan, Popa,
  Wittmann, Neumann, Jelezko, Rabeau, Stavrias, Greentree, Prawer
  et~al.}}]{GaebelNature2006}
\bibinfo{author}{\bibfnamefont{T.}~\bibnamefont{Gaebel}},
  \bibinfo{author}{\bibfnamefont{M.}~\bibnamefont{Domhan}},
  \bibinfo{author}{\bibfnamefont{I.}~\bibnamefont{Popa}},
  \bibinfo{author}{\bibfnamefont{C.}~\bibnamefont{Wittmann}},
  \bibinfo{author}{\bibfnamefont{P.}~\bibnamefont{Neumann}},
  \bibinfo{author}{\bibfnamefont{F.}~\bibnamefont{Jelezko}},
  \bibinfo{author}{\bibfnamefont{J.~R.} \bibnamefont{Rabeau}},
  \bibinfo{author}{\bibfnamefont{N.}~\bibnamefont{Stavrias}},
  \bibinfo{author}{\bibfnamefont{A.~D.} \bibnamefont{Greentree}},
  \bibinfo{author}{\bibfnamefont{S.}~\bibnamefont{Prawer}},
  \bibnamefont{et~al.}, \bibinfo{journal}{Nat. Phys.}
  \textbf{\bibinfo{volume}{2}}, \bibinfo{pages}{408} (\bibinfo{year}{2006}).

\bibitem[{\citenamefont{Takahashi et~al.}(2008)\citenamefont{Takahashi, Hanson,
  van Tol, Sherwin, and Awschalom}}]{TakahashiPRL2008}
\bibinfo{author}{\bibfnamefont{S.}~\bibnamefont{Takahashi}},
  \bibinfo{author}{\bibfnamefont{R.}~\bibnamefont{Hanson}},
  \bibinfo{author}{\bibfnamefont{J.}~\bibnamefont{van Tol}},
  \bibinfo{author}{\bibfnamefont{M.~S.} \bibnamefont{Sherwin}},
  \bibnamefont{and} \bibinfo{author}{\bibfnamefont{D.~D.}
  \bibnamefont{Awschalom}}, \bibinfo{journal}{Phys. Rev. Lett.}
  \textbf{\bibinfo{volume}{101}}, \bibinfo{pages}{047601}
  (\bibinfo{year}{2008}).

\bibitem[{\citenamefont{Balasubramanian
  et~al.}(2009)\citenamefont{Balasubramanian, Neumann, Twitchen, Markham,
  Kolesov, Mizuochi, Isoya, Achard, Beck, Tissler
  et~al.}}]{BalasubramanianNatMater2009}
\bibinfo{author}{\bibfnamefont{G.}~\bibnamefont{Balasubramanian}},
  \bibinfo{author}{\bibfnamefont{P.}~\bibnamefont{Neumann}},
  \bibinfo{author}{\bibfnamefont{D.}~\bibnamefont{Twitchen}},
  \bibinfo{author}{\bibfnamefont{M.}~\bibnamefont{Markham}},
  \bibinfo{author}{\bibfnamefont{R.}~\bibnamefont{Kolesov}},
  \bibinfo{author}{\bibfnamefont{N.}~\bibnamefont{Mizuochi}},
  \bibinfo{author}{\bibfnamefont{J.}~\bibnamefont{Isoya}},
  \bibinfo{author}{\bibfnamefont{J.}~\bibnamefont{Achard}},
  \bibinfo{author}{\bibfnamefont{J.}~\bibnamefont{Beck}},
  \bibinfo{author}{\bibfnamefont{J.}~\bibnamefont{Tissler}},
  \bibnamefont{et~al.}, \bibinfo{journal}{Nat. Mater.}
  \textbf{\bibinfo{volume}{8}}, \bibinfo{pages}{383} (\bibinfo{year}{2009}).

\bibitem[{\citenamefont{de~Lange et~al.}(2010)\citenamefont{de~Lange, Wang,
  Riste, Dobrovitski, and Hanson}}]{LangeScience2010}
\bibinfo{author}{\bibfnamefont{G.}~\bibnamefont{de~Lange}},
  \bibinfo{author}{\bibfnamefont{Z.~H.} \bibnamefont{Wang}},
  \bibinfo{author}{\bibfnamefont{D.}~\bibnamefont{Riste}},
  \bibinfo{author}{\bibfnamefont{V.~V.} \bibnamefont{Dobrovitski}},
  \bibnamefont{and} \bibinfo{author}{\bibfnamefont{R.}~\bibnamefont{Hanson}},
  \bibinfo{journal}{Science} \textbf{\bibinfo{volume}{330}},
  \bibinfo{pages}{60} (\bibinfo{year}{2010}).

\bibitem[{\citenamefont{Bar-Gill et~al.}(2013)\citenamefont{Bar-Gill, Pham,
  Jarmola, Budker, and Walsworth}}]{BarGillNatCommun2013}
\bibinfo{author}{\bibfnamefont{N.}~\bibnamefont{Bar-Gill}},
  \bibinfo{author}{\bibfnamefont{L.~M.} \bibnamefont{Pham}},
  \bibinfo{author}{\bibfnamefont{A.}~\bibnamefont{Jarmola}},
  \bibinfo{author}{\bibfnamefont{D.}~\bibnamefont{Budker}}, \bibnamefont{and}
  \bibinfo{author}{\bibfnamefont{R.~L.} \bibnamefont{Walsworth}},
  \bibinfo{journal}{Nat. Commun.} \textbf{\bibinfo{volume}{4}},
  \bibinfo{pages}{1743} (\bibinfo{year}{2013}).

\end{thebibliography}

\section*{Acknowledgements}

We acknowledge Professor I. Cirac for the inspiring discussions on the
physical understanding the scaling relation. We thank Professor H.-D. Yuan,
S.-L. Luo, X.-G. Wang, and Doctor Y. Yao for fruitful suggestions and comments
on the manuscript. N.Z. is supported by NKBRP (973 Program) 2014CB848700 and
NSFC Nos. 11374032 and 11121403. W.Y. is supported by NSFC Nos. 11774021,
11274036, and 11322542. C.P.S. is supported by NSFC Nos. 11421063, 11534002,
11121403, the national key research and development program (Grant No.
2016YFA0301201), and the National 973 program (Grants No. 2012CB922104 and No.
2014CB921403). We acknowledge support by NSFC program for 'Scientific Research
Center' (Program No. U1530401).

\section*{Author contributions statement}

N. Z. conceived the idea, Y. N. F. formulated the theories for vacuum initial
state of the oscillator and Carr-Purcell-Meiboom-Gill control on the
oscillator, W. Y. generalized the theories to arbitrary initial states and
arbitrary periodic quantum controls. N. Z. and Y. N. F. wrote the first
version of the paper. W. Y. wrote the final version. All authors discussed the
results and commented on the manuscript.

\section*{Additional information}

Competing financial interests: The authors declare no competing financial interests.
\appendix 
Here we describe the adaptive quantum control scheme for quantum sensing and
analyze its performance. Two kinds of resources can be utilized to improve the
precision: repeated measurements (as quantified by the number $\nu$ of
repetition) is a classical resource that improves the precision according to
the classical scaling $\delta\omega\propto1/\sqrt{\nu}$; while the evolution
time $T$ is a quantum resource that improves the precision according to the
quantum enhanced scaling $\delta\omega\propto1/T^{2}$. When the total resource
-- the total time cost $\mathbb{T}$ -- is fixed, it is desirable to spend more
resources on $T$ instead of $\nu$. An extreme case is to spend all the time
cost on the quantum resource, i.e., a single measurement $(\nu=1$) with the
evolution time $T=\mathbb{T}$.

\section{Adaptive quantum control:\ analytical analysis}

Recall that when
\begin{equation}
\sqrt{2\bar{n}+1}\left\vert \alpha\right\vert \ll1, \label{CD2}%
\end{equation}
we obtain the sensing precision
\[
\delta\omega\approx\frac{\pi}{g(\zeta)\tilde{\lambda}T^{2}},
\]
where $\tilde{\lambda}\equiv\sqrt{2\bar{n}+1}\left\vert \alpha_{1}\right\vert
/\tau$ is nearly a constant and
\[
g(\zeta)\approx\left\vert \frac{\pi\zeta\cos(\pi\zeta)-\sin(\pi\zeta)}%
{\pi\zeta^{2}}\right\vert
\]
is a function of $\zeta\equiv N\left(  \omega\tau/2\pi-1\right)  $. Ideally,
we should first set $\tau=(2\pi/\omega)(1+1/N)$ to make $\zeta=1$ and then
increase $N$ to increase $T\equiv N\tau$. Setting $\zeta=1$ exactly not only
makes $|\alpha|=0$ to satisfy Eq. (\ref{CD2}), but also makes $g=1$ to achieve
the sensing precision%
\begin{equation}
\delta\omega(T)\approx\frac{\pi}{\tilde{\lambda}T^{2}}. \label{DW2}%
\end{equation}
However, if our knowledge about $\omega$ has an uncertainty $\delta\omega$,
then we suffer from an uncertainty $\delta\zeta\equiv(T/2\pi)\delta\omega$ in
tuning the value of $\zeta$, i.e., we cannot set $\zeta=1$ exactly, but
instead only make $\zeta\in\lbrack1-\delta\zeta,1+\delta\zeta]$. In this case,
the actual sensing precision is roughly given by
\begin{equation}
\delta\omega_{\mathrm{i}}(T)\approx\dfrac{\pi}{g_{\mathrm{rms}}\tilde{\lambda
}T^{2}}, \label{DW1}%
\end{equation}
where $g_{\mathrm{rms}}\equiv\sqrt{\langle g^{2}\rangle}$ and $\langle
g^{2}\rangle$ is the average of $g^{2}(\zeta)$ over the region $[1-\delta
\zeta,1+\delta\zeta]$. Since $\langle g^{2}\rangle\sim1$ when $\delta
\zeta\lesssim1$ but $\langle g^{2}\rangle\propto1/\delta\zeta$ when
$\delta\zeta\gg1$, to achieve the $1/T^{2}$ scaling, we should ensure both Eq.
(\ref{CD2}) and
\begin{equation}
\delta\zeta\lesssim1. \label{CD1}%
\end{equation}
In the following, we assume $\tilde{\lambda}\ll\omega$, which is typically the
case in hybrid quantum systems.

In early stages of the sensing (i.e., $\delta\omega\gg\tilde{\lambda}$), Eq.
(\ref{CD1}) limits the coherent evolution time to $T\lesssim2\pi/\delta
\omega\ll2\pi/\tilde{\lambda}$. Then, using $|K|\leq N$ gives $\sqrt{2\bar
{n}+1}\left\vert \alpha\right\vert \leq\tilde{\lambda}T$, so Eq. (\ref{CD2})
is satisfied automatically. Therefore, in the early stages of the sensing, we
need only satisfy Eq. (\ref{CD1}) by setting%
\begin{equation}
T\approx\dfrac{1}{\kappa_{\mathrm{i}}}\dfrac{2\pi}{\delta\omega},
\label{CD1_T}%
\end{equation}
where $\kappa_{\mathrm{i}}\gtrsim1$. In this case, we have $\delta\zeta
\approx1/\kappa_{\mathrm{i}}\lesssim1$, so the sensing precision is given by
Eq. (\ref{DW1}).

As the sensing goes on, $\delta\omega$ becomes smaller than $\tilde{\lambda}$,
then using $\tau\approx2\pi/\omega$, we have $N\approx\omega T/(2\pi)\gg1,$ so
$\sqrt{2\bar{n}+1}|\alpha|\approx\tilde{\lambda}T\left\vert \sin(\pi
\zeta)/(\pi\zeta)\right\vert \sim\tilde{\lambda}T\delta\zeta$, so Eq.
(\ref{CD2}) amounts to%
\[
\delta\zeta\ll\frac{1}{\tilde{\lambda}T}\Leftrightarrow T\ll\sqrt{\frac{2\pi
}{\tilde{\lambda}\delta\omega}}.
\]
To satisfy Eqs. (\ref{CD2}) and (\ref{CD1}) simultaneously, we set%
\begin{equation}
T\approx\dfrac{1}{\kappa}\sqrt{\dfrac{2\pi}{\tilde{\lambda}\delta\omega}},
\label{CD2_T}%
\end{equation}
where $\kappa\gg1$. Under this condition, we have $\delta\zeta=(1/\kappa
)\sqrt{\delta\omega/(2\pi\tilde{\lambda})}\ll1$, so the sensing precision is
given by Eq. (\ref{DW2}).

Next we describe the adaptive quantum sensing schemes capable of extending the
$1/T^{2}$ scaling to arbitrarily long $T$. Before the quantum sensing, our
prior knowledge about $\omega$ is quantified by a Gaussian distribution%
\begin{equation}
P_{0}(\omega)=\frac{1}{\sqrt{2\pi}\delta\omega_{0}}e^{-(\omega-\omega_{0}%
)^{2}/[2(\delta\omega_{0})^{2}]}, \label{P0W}%
\end{equation}
corresponding to an unbiased estimator $\omega_{0}$ with a precision (or
uncertainty) $\tilde{\lambda}\ll\delta\omega_{0}\ll\omega$. The adaptive
scheme consists of many steps. The central idea is to utilize the measurements
in each step to successively refine our knowledge about $\omega$ and reduce
the uncertainty $\delta\omega$, so that we can use successively longer
coherent evolution time in the next step. The entire adaptive scheme consists
of two stages: (i) $\delta\omega\gtrsim\tilde{\lambda}$, where we choose $T$
according to Eq. (\ref{CD1_T}) to achieve Eq. (\ref{DW1}); and (ii)
$\delta\omega\lesssim\tilde{\lambda}$, where we choose $T$ according to Eq.
(\ref{CD2_T}) to achieve Eq. (\ref{DW2}).

\subsection{Stage (i): $\delta\omega\gtrsim\tilde{\lambda}$}

In this stage, the large uncertainty $\delta\omega$ only allows short
evolution times, so a single measurement only improves the precision slightly.
Therefore, we need to utilize the classical resources (i.e., repeated
measurements) to boost the improvement of the precision:

\textbf{Step 1.} We require the pulse interval $\tau_{1}$ and the pulse number
$N_{1}$ to satisfy $\omega_{0}\tau_{1}-2\pi=2\pi/N_{1}$ and the evolution time
$T_{1}\equiv N_{1}\tau_{1}$ to be close to $(1/\kappa_{\mathrm{i}})2\pi
/\delta\omega_{0}$, where $\kappa_{\mathrm{i}}\gtrsim1$ is a constant
parameter. Then we repeat the projective $\hat{\sigma}_{x}$ measurements on
the qubit for $\nu_{1}$ times and obtain the measurement outcomes
$\mathbf{u}_{1}\equiv(u_{1},u_{2},\cdots,u_{\nu_{1}})$. Next we combine our
prior knowledge and the new information from the outcomes $\mathbf{u}_{1}$ to
update the distribution for $\omega$ from $P_{0}(\omega)$ to
\[
P_{\mathbf{u}_{1}}(\omega)=\frac{P_{0}(\omega)P(\mathbf{u}_{1}|\omega)}{\int
P_{0}(\omega)P(\mathbf{u}_{1}|\omega)d\omega},
\]
where $P(\mathbf{u}_{1}|\omega)$ is the probability for obtaining the outcome
$\mathbf{u}_{1}$. Then we construct the maximum likelihood estimator
\[
\omega_{1}=\arg\max_{\omega}P_{\mathbf{u}_{1}}(\omega)
\]
as the position of the maximum of $P_{\mathbf{u}_{1}}(\omega)$. For large
$\nu_{1}$, the maximum likelihood estimator attains the Cram\'{e}r-Rao bound,
so its precision (or uncertainty) $\delta\omega_{1}$ is estimated by using the
Cram\'{e}r-Rao bound as
\[
\delta\omega_{1}=\frac{1}{\sqrt{(\delta\omega_{0})^{-2}+\nu_{1}[\delta
\omega_{\mathrm{i}}(T_{1})]^{-2}}}=\frac{\delta\omega_{0}}{\sqrt{1+\nu
_{1}G_{1}^{2}}},
\]
where%
\[
G_{1}\equiv\frac{\delta\omega_{0}}{\delta\omega_{\mathrm{i}}(T_{1})}%
\approx\eta_{\mathrm{i}}\frac{\tilde{\lambda}}{\delta\omega_{0}}%
\]
quantifies the information gain $\delta\omega_{\mathrm{i}}(T_{1})$ [Eq.
(\ref{DW1})] from a single measurement relative to the prior knowledge
$\delta\omega_{0}$ and%
\begin{equation}
\eta_{\mathrm{i}}\equiv\frac{4\pi g_{\mathrm{rms}}}{\kappa_{\mathrm{i}}^{2}%
}\sim1.\label{YITA}%
\end{equation}
Initially $\delta\omega_{0}\gg\tilde{\lambda}$, so $G_{1}\ll1$, i.e., a single
measurement only improves the precision slightly. Then we have to utilize the
classical resource $\nu_{1}\gg1$ to boost the improvement of the precision.
Taking $\nu_{1}=c_{\mathrm{i}}^{2}/G_{1}^{2}$ ($c_{\mathrm{i}}$ is a constant
parameter) improves the precision by a factor $\sqrt{1+c_{\mathrm{i}}^{2}}$:%
\[
\delta\omega_{1}\approx\frac{\delta\omega_{0}}{\sqrt{1+c_{\mathrm{i}}^{2}}}.
\]
The time cost of this step is
\[
\nu_{1}T_{1}\approx\frac{c_{\mathrm{i}}^{2}\kappa_{\mathrm{i}}^{3}}{8\pi
g_{\mathrm{rms}}^{2}}\frac{\delta\omega_{0}}{\tilde{\lambda}^{2}}.
\]

\textbf{Step 2.} We require the pulse interval $\tau_{2}$ and the pulse number
$N_{2}$ to satisfy $\omega_{1}\tau_{2}-2\pi=2\pi/N_{2}$ and the evolution time
$T_{2}\equiv N_{2}\tau_{2}$ to be close to $(1/\kappa_{\mathrm{i}})2\pi
/\delta\omega_{1}\approx\sqrt{1+c_{\mathrm{i}}^{2}}T_{1}$. Then we repeat the
projective $\hat{\sigma}_{x}$ measurement on the qubit for $\nu_{2}$ times and
obtain the measurement outcomes $\mathbf{u}_{2}\equiv(u_{1},u_{2}%
,\cdots,u_{\nu_{2}})$. Next we combine our previous knowledge $P_{\mathbf{u}%
_{1}}(\omega)$ and the new information from the outcomes $\mathbf{u}_{2}$ to
update the distribution for $\omega$ to
\[
P_{\mathbf{u}_{1}\mathbf{u}_{2}}(\omega)=\frac{P_{\mathbf{u}_{1}}%
(\omega)P(\mathbf{u}_{2}|\omega)}{\int P_{\mathbf{u}_{1}}(\omega
)P(\mathbf{u}_{2}|\omega)d\omega},
\]
where $P(\mathbf{u}_{2}|\omega)$ is the probability for obtaining the outcome
$\mathbf{u}_{2}$. Then we construct the maximum likelihood estimator
$\omega_{2}$ as the position of the maximum of the probability distribution
$P_{\mathbf{u}_{1}\mathbf{u}_{2}}(\omega)$. The precision (or uncertainty)
$\delta\omega_{2}$ is estimated by the Cram\'{e}r-Rao bound as%
\[
\delta\omega_{2}\approx\frac{1}{\sqrt{(\delta\omega_{1})^{-2}+\nu_{2}%
[\delta\omega_{\mathrm{i}}(T_{2})]^{-2}}}=\frac{\delta\omega_{1}}{\sqrt
{1+\nu_{2}G_{2}^{2}}},
\]
where the relative information gain
\[
G_{2}\equiv\frac{\delta\omega_{1}}{\delta\omega_{\mathrm{i}}(T_{2})}%
\approx\eta_{\mathrm{i}}\frac{\tilde{\lambda}}{\delta\omega_{1}}\approx
\sqrt{1+c_{\mathrm{i}}^{2}}G_{1}%
\]
is larger than the previous step due to the longer evolution time. Thus we
need only utilize less classical resources $\nu_{2}=c_{\mathrm{i}}^{2}%
/G_{2}^{2}\approx\nu_{1}/(1+c_{\mathrm{i}}^{2})$ to improve the precision by
the same factor $\sqrt{1+c_{\mathrm{i}}^{2}}$:%
\[
\delta\omega_{2}\approx\frac{\delta\omega_{1}}{\sqrt{1+c_{\mathrm{i}}^{2}}}.
\]
The time cost of this step is $\nu_{2}T_{2}\approx\nu_{1}T_{1}/\sqrt
{1+c_{\mathrm{i}}^{2}}$.

\textbf{Step }$m$\textbf{. }We require the pulse interval $\tau_{m}$ and the
pulse number $N_{m}$ to satisfy $\omega_{m-1}\tau_{m}-2\pi=2\pi/N_{m}$ and the
evolution time $T_{m}\equiv N_{m}\tau_{m}$ to be close to $(1/\kappa
_{\mathrm{i}})2\pi/\delta\omega_{m-1}\approx\sqrt{1+c_{\mathrm{i}}^{2}}%
T_{m-1}$. Then we repeat the projective $\hat{\sigma}_{x}$ measurement on the
qubit for $\nu_{m}$ times to obtain the maximum likelihood estimator
$\omega_{m}$, whose precision is estimated as%
\[
\delta\omega_{m}\approx\frac{\delta\omega_{m-1}}{\sqrt{1+\nu_{m}G_{m}^{2}}},
\]
where the relative information gain
\[
G_{m}\equiv\frac{\delta\omega_{m-1}}{\delta\omega_{\mathrm{i}}(T_{m})}%
\approx\eta_{\mathrm{i}}\frac{\tilde{\lambda}}{\delta\omega_{m-1}}\approx
\sqrt{1+c_{\mathrm{i}}^{2}}G_{m-1}.
\]
As long as $\delta\omega_{m-1}\gg\tilde{\lambda}$, we have $G_{m}\ll1$, so we
still need to utilize the classical resource $\nu_{m}=c_{\mathrm{i}}^{2}%
/G_{m}^{2}\approx\nu_{m-1}/(1+c_{\mathrm{i}}^{2})$ to boost the improvement of
the precision by a factor $\sqrt{1+c_{\mathrm{i}}^{2}}$:%
\[
\delta\omega_{m}\approx\frac{\delta\omega_{m-1}}{\sqrt{1+c_{\mathrm{i}}^{2}}}.
\]
The time cost of this step is $\nu_{m}T_{m}\approx\nu_{m-1}T_{m-1}%
/\sqrt{1+c_{\mathrm{i}}^{2}}$.

This stage stops when the precision $\delta\omega$ becomes comparable or less
than $\tilde{\lambda}$, so that a single measurement can lead to significant
precision improvement.

In this stage, we have introduced two constant parameters $\kappa_{\mathrm{i}%
}$ and $c_{\mathrm{i}}$: the former ensures Eq. (\ref{CD1}) is satisfied in
every step, while the latter quantifies the classical resource to be utilized
in each step. Every step improves the precision by a factor of $\sqrt
{1+c_{\mathrm{i}}^{2}}$, but the time cost is $1/\sqrt{1+c_{\mathrm{i}}^{2}}$
times that of the previous step, consistent with the $1/T^{2}$ scaling of the
sensing precision. The case $c_{\mathrm{i}}\gg1$ corresponds to significant
improvement of the precision in each step ($\delta\omega_{m}\ll\delta
\omega_{m-1}$), so that the evolution time of the next step can be prolonged
significantly ($T_{m}\gg T_{m-1}$); while $c_{\mathrm{i}}\ll1$ corresponds to
small improvement of the precision in each step ($\delta\omega_{m}%
\lesssim\delta\omega_{m-1}$), so that the evolution time of the next step can
only be prolonged slightly ($T_{m}\gtrsim T_{m-1}$).

At the end of the $m$th step, the time cost is%
\[
T_{\mathrm{i}}\equiv\nu_{1}T_{1}+\cdots+\nu_{m}T_{m}\approx\nu_{1}T_{1}%
\frac{1-\frac{1}{(\sqrt{1+c_{\mathrm{i}}^{2}})^{m}}}{1-\frac{1}{\sqrt
{1+c_{\mathrm{i}}^{2}}}}%
\]
and the precision is%
\[
\delta\omega_{m}\approx\frac{\delta\omega_{0}}{(\sqrt{1+c_{\mathrm{i}}^{2}%
})^{m}}.
\]
For $c_{\mathrm{i}}\ll1$ but large $m$ so that the overall precision
improvement is significant, i.e., $(\sqrt{1+c_{\mathrm{i}}^{2}})^{m}\gg1$, the
time cost%
\[
T_{\mathrm{i}}\approx\frac{2\nu_{1}T_{1}}{c_{\mathrm{i}}^{2}}=\frac
{\kappa_{\mathrm{i}}^{3}}{4\pi g_{\mathrm{rms}}^{2}}\frac{\delta\omega_{0}%
}{\tilde{\lambda}^{2}}%
\]
is independent of $c_{\mathrm{i}}$ and the number of steps $m$. When
$c_{\mathrm{i}}\gg1$, the time cost is dominated by the first step:$\ $%
\[
T_{\mathrm{i}}\approx\nu_{1}T_{1}\approx\frac{c_{\mathrm{i}}^{2}}{2}%
\frac{\kappa_{\mathrm{i}}^{3}}{4\pi g_{\mathrm{rms}}^{2}}\frac{\delta
\omega_{0}}{\tilde{\lambda}^{2}}%
\]
and is still independent of $m$. The case $c_{\mathrm{i}}\ll1$ requires less
time cost than the case $c_{\mathrm{i}}\gg1$, because the latter utilizes more
classical resources (i.e., repeated measurements). On the other hand, in order
to improve the precision from $\delta\omega_{0}$ to the desired precision
$\tilde{\lambda}$, the case $c_{\mathrm{i}}\ll1$ requires much more adaptive
steps than the case $c_{\mathrm{i}}\gg1$, because when $c_{\mathrm{i}}\ll1$
($c_{\mathrm{i}}\gg1$), the precision is improved slightly (significantly) in
each step.

\subsection{Stage (ii): $\delta\omega\lesssim\tilde{\lambda}$}

At the beginning of this stage, we have an estimator $\omega_{\mathrm{i}}$
(i.e., the estimator at the end of the previous stage) with a precision
$\delta\omega_{\mathrm{i}}\sim\tilde{\lambda}$. In this stage, the small
uncertainty $\delta\omega$ allows long evolution time so that a single
measurement may significantly improve the precision.

\textbf{Step 1. }We require the pulse interval $\tau_{1}$ and the pulse number
$N_{1}$ to satisfy $\omega_{\mathrm{i}}\tau_{1}-2\pi=2\pi/N_{1}$ and the
evolution time $T_{1}\equiv N_{1}\tau_{1}$ to be close to $(1/\kappa
)\sqrt{2\pi/(\tilde{\lambda}\delta\omega_{\mathrm{i}})}$, where $\kappa\gg1$
is a constant parameter. Then we repeat the projective $\hat{\sigma}_{x}$
measurements on the qubit for $\nu$ times and construct the maximum likelihood
estimator $\omega_{1}$. The precision of $\omega_{1}$ is estimated as
\[
\delta\omega_{1}\approx\frac{\delta\omega_{\mathrm{i}}}{\sqrt{1+c^{2}}},
\]
where $c\equiv\sqrt{\nu}\eta$,
\begin{equation}
\eta\equiv\frac{\delta\omega_{\mathrm{i}}}{\delta\omega(T_{1})}\approx\frac
{2}{\kappa^{2}} \label{YITA2}%
\end{equation}
quantifies the relative information gain from a single measurement, and
$\delta\omega(T_{1})$ is given by Eq. (\ref{DW2}).

\textbf{Step 2.} We require the pulse interval $\tau_{2}$ and the pulse number
$N_{2}$ to satisfy $\omega_{1}\tau_{2}-2\pi=2\pi/N_{2}$ and the evolution time
$T_{2}\equiv N_{2}\tau_{2}$ to be close to $(1/\kappa)\sqrt{2\pi
/(\tilde{\lambda}\delta\omega_{1})}\approx(1+c^{2})^{1/4}T_{1}$. Then we
repeat the projective $\hat{\sigma}_{x}$ measurement on the qubit for $\nu$
times to obtain the maximum likelihood estimator $\omega_{2}$, whose precision
is estimated as%
\[
\delta\omega_{2}\approx\frac{\delta\omega_{1}}{\sqrt{1+c^{2}}},
\]
where we have used $\delta\omega_{1}/\delta\omega(T_{2})\approx\eta$.

\textbf{Step }$m$\textbf{. }We require the pulse interval $\tau_{2}$ and the
pulse number $N_{2}$ to satisfy $\omega_{m-1}\tau_{m}-2\pi=2\pi/N_{m}$ and the
evolution time $T_{m}\equiv N_{m}\tau_{m}$ to be close to $(1/\kappa
)\sqrt{2\pi/(\tilde{\lambda}\delta\omega_{m-1})}\approx(1+c^{2})^{1/4}T_{m-1}%
$. Then we repeat the projective $\hat{\sigma}_{x}$ measurement on the qubit
for $\nu$ times to obtain the maximum likelihood estimator $\omega_{m}$, whose
precision is estimated as%
\[
\delta\omega_{m}\approx\frac{\delta\omega_{m-1}}{\sqrt{1+c^{2}}},
\]
where we have used $\delta\omega_{m-1}/\delta\omega(T_{m})\approx\eta$.

In this stage, we have introduced two parameters $\kappa$ and $c$: the former
ensures Eq. (\ref{CD2}) is satisfied in every step, while the latter
quantifies the classical resource to be utilized in each step. Every step
improves the precision by a factor of $\sqrt{1+c^{2}}$ and uses a time cost
that is $(1+c^{2})^{1/4}$ times that of the previous step, consistent with the
$1/T^{2}$ scaling of the sensing precision. The case $c\gg1$ corresponds to
significant improvement of the precision in each step ($\delta\omega_{m}%
\ll\delta\omega_{m-1}$), so that the evolution time of the next step can be
prolonged significantly ($T_{m}\gg T_{m-1}$); while the case $c\ll1$
corresponds to small improvement of the precision in each step ($\delta
\omega_{m}\lesssim\delta\omega_{m-1}$), so that the evolution time of the next
step can only be prolonged slightly ($T_{m}\gtrsim T_{m-1}$).

At the end of the $m$th step, the time cost is
\[
T_{\mathrm{ii}}=\nu(T_{1}+\cdots+T_{m})\approx\nu T_{1}\frac{(1+c^{2}%
)^{m/4}-1}{(1+c^{2})^{1/4}-1},
\]
and the final precision is%
\[
\delta\omega_{m}\approx\frac{\delta\omega_{\mathrm{i}}}{(\sqrt{1+c^{2}})^{m}}.
\]
For $c\ll1$, we have%
\[
\delta\omega_{m}\approx\frac{16}{\eta^{3}}\frac{\pi}{\tilde{\lambda
}T_{\mathrm{ii}}^{2}}\approx2\kappa^{6}\frac{\pi}{\tilde{\lambda
}T_{\mathrm{ii}}^{2}}.
\]
For $c\gg1$, the total time cost is dominated by the last step:
$T_{\mathrm{ii}}\approx\nu T_{m}$. The final precision is also dominated by
the last step:
\begin{equation}
\delta\omega_{m}\approx\frac{\delta\omega(T_{m})}{\sqrt{\nu}}\approx\nu
^{3/2}\frac{\pi}{\tilde{\lambda}T_{\mathrm{ii}}^{2}},\label{DWII}%
\end{equation}
where $\delta\omega(T)$ is given in Eq. (\ref{DW2}). Obviously, the case
$c\ll1$ provides better sensing precision than $c\gg1$.

\section{Adaptive quantum control:\ numerical implementation}

In our numerical simulation, we consider the $N$-period
Carr--Purcell--Meiboom--Gill (CPMG) sequence consisting of $N$ identical
control units $\tau/4$-$\pi$-$\tau/2$-$\pi$-$\tau/4$, corresponding to
\[
\alpha_{1}(\omega,\tau)=i\frac{8\lambda}{\omega}e^{i\omega\tau/2}\cos
\frac{\omega\tau}{8}\sin^{3}\frac{\omega\tau}{8}%
\]
and hence%
\[
\alpha(N,\omega,\tau)=\alpha_{1}(\omega,\tau)\sum_{n=0}^{N-1}e^{in\omega\tau
}.
\]
The initial state of the harmonic oscillator is taken as the thermal state
$\rho=e^{-\omega a^{\dagger}a/(k_{B}T)}/\operatorname*{Tr}e^{-\omega
a^{\dagger}a/(k_{B}T)}$, as characterized by the thermal population $\bar
{n}=1/(e^{\omega/(k_{B}T)}-1)$. In this case, the off-diagonal coherence of
the qubit is $L=e^{-2(2\bar{n}+1)|\alpha|^{2}}$ and the probability
distribution of the $\sigma_{x}$ measurement is
\[
P(\pm1|\omega)=\frac{1\pm e^{-2(2\bar{n}+1)|\alpha|^{2}}}{2}.
\]

\subsection{Stage (i)}

The input/control parameters include $\kappa_{\mathrm{i}}$, $c_{\mathrm{i}}$,
$\bar{n}$, and the prior distribution $P_{0}(\omega)$ [Eq. (\ref{P0W})] for
the unknown frequency $\omega$, as characterized by an estimator $\omega_{0}$
and its uncertainty $\delta\omega_{0}$.

At the beginning of the $k$-th adaptive step, we already have a probability
distribution $P_{k-1}(\omega)$ from the previous steps, which gives an
estimator $\omega_{k-1}$ and its uncertainty $\delta\omega_{k-1}$. In the
$k$-th step, we apply the CPMG sequence with $N_{k}$ identical control units
$\tau_{k}/4$-$\pi$-$\tau_{k}/2$-$\pi$-$\tau_{k}/4$ and repeat the measurements
for $\nu_{k}$ times, where
\begin{align}
N_{k}  &  =\mathrm{nint}(\frac{\omega_{k-1}}{\kappa_{\mathrm{i}}\delta
\omega_{k-1}}-1),\\
\tau_{k}  &  =\frac{2\pi}{\omega_{k-1}}(1+\frac{1}{N_{k}}),\\
\nu_{k}  &  =\max\{\mathrm{nint}\frac{c_{\mathrm{i}}^{2}(\delta\omega
_{k-1})^{2}}{\tilde{\lambda}_{k}^{2}\eta_{\mathrm{i}}^{2}},1\},
\end{align}
with $\mathrm{nint}(a)$ for the integer closest to $a$, $\tilde{\lambda}%
_{k}\equiv\sqrt{2\bar{n}+1}|\alpha_{1}(\tau_{k},\omega_{k-1})|/\tau_{k}$,
$\eta_{\mathrm{i}}$ given by Eq. (\ref{YITA}), and $g_{\mathrm{rms}}%
\approx0.83544$ is obtained by taking $\delta\zeta=1$. Next, we calculate
$\alpha_{k}=\alpha(N_{k},\omega,\tau_{k})$ and $P_{k}(\pm1|\omega)=(1\pm
e^{-2(2\bar{n}+1)|\alpha_{k}|^{2}})/2$, randomly generate $\nu_{k}$ outcomes
according to $P_{k}(\pm1|\omega)$, and use $N_{\pm}$ to denote the number of
outcome $\pm1$ in those $\nu_{k}$ results. Then we calculate the updated
probability distribution function
\[
P_{k}(\omega)=P_{k-1}(\omega)[P_{k}(+1|\omega)]^{N_{+}}[P_{k}(-1|\omega
)]^{N_{-}}%
\]
and obtain the maximum likelihood estimator $\omega_{k}\equiv\arg\max_{\omega
}P_{k}(\omega)$ as the location of the maximum of $P_{k}(\omega)$ as a
function of $\omega$. Finally, we calculate the uncertainty of $\omega_{k}$
by
\[
\delta\omega_{k}=\left[  \frac{\int d\omega(\omega_{k}-\omega)^{2}P_{k}%
(\omega)}{\int d\omega P_{k}(\omega)}\right]  ^{\frac{1}{2}}.
\]

When $\delta\omega_{k}<\tilde{\lambda}_{k}$, this stage stops and we begin
stage (ii) with
\begin{align*}
\omega_{\mathrm{i}}  &  =\omega_{k},\\
\delta\omega_{\mathrm{i}}  &  =\delta\omega_{k},\\
P_{\mathrm{i}}(\omega)  &  =P_{k}(\omega).
\end{align*}

\subsection{Stage (ii)}

The input/control parameters include $\kappa$, $c$, $\bar{n}$, and the
distribution $P_{\mathrm{i}}(\omega)$, as characterized by an estimator
$\omega_{\mathrm{i}}$ and its uncertainty $\delta\omega_{\mathrm{i}}$. At the
beginning of the $k$-th adaptive step, we already have a probability
distribution $P_{k-1}(\omega)$ from the previous steps, which gives an
estimator $\omega_{k-1}$ and uncertainty $\delta\omega_{k-1}$. In the $k$-th
adaptive step, we apply the CPMG sequence with $N_{k}$ identical control units
$\tau_{k}/4$-$\pi$-$\tau_{k}/2$-$\pi$-$\tau_{k}/4$ and repeat the measurements
for $\nu$ times, where%
\begin{align*}
N_{k} &  =\mathrm{nint}(\frac{\omega_{k-1}}{\kappa\sqrt{2\pi\tilde{\lambda
}\delta\omega_{k-1}}}-1),\\
\tau_{k} &  =\frac{2\pi}{\omega_{k-1}}(1+\frac{1}{N_{k}}),\\
\nu &  =\frac{c^{2}\kappa^{4}}{4},
\end{align*}
and $\tilde{\lambda}=\lambda\sqrt{2\bar{n}+1}/\pi$. Next, we calculate
$\alpha_{k}=\alpha(N_{k},\omega,\tau_{k})$ and $P_{k}(\pm1|\omega)=(1\pm
e^{-2(2\bar{n}+1)|\alpha_{k}|^{2}})/2$. Then we randomly generate $\nu$
outcomes according to $P_{k}(\pm1|\omega)$, and let $N_{\pm}$ denote the
number of outcome $\pm1$ in those $\nu$ outcomes. Then we calculate the
updated probability distribution function
\[
P_{k}(\omega)=P_{k-1}(\omega)[P_{k}(+1|\omega)]^{N_{+}}[P_{k}(-1|\omega
)]^{N_{-}}%
\]
and obtain the maximum likelihood estimator $\omega_{k}\equiv\arg\max_{\omega
}P_{k}(\omega)$. Finally, we calculate the uncertainty of $\omega_{k}$ by
\[
\delta\omega_{k}=\left[  \frac{\int d\omega(\omega_{k}-\omega)^{2}P_{k}%
(\omega)}{\int d\omega P_{k}(\omega)}\right]  ^{\frac{1}{2}}.
\]
This process can be continued until the uncertainty $\delta\omega_{k}$ reaches
the desired precision.

In the numerical simulation, we take $\bar{n}=10$ and $\bar{n}=1000$,
respectively, $\lambda=0.1$, $\delta\omega_{0}=0.5$, $\omega_{0}=50.5$,
$\omega=50$, $c_{\mathrm{i}}=c=0.1$ and $\kappa_{\mathrm{i}}=\kappa=2.$ The
total time cost is $\mathbb{T}=T_{\mathrm{i}}+T_{\mathrm{ii}}$.

\end{document}